\newif\ifsingle
\singletrue 

\newif\iflowsnr
\lowsnrtrue 

\newif\ifDFTcomment

\ifsingle
\documentclass[11pt,draftcls, onecolumn]{IEEEtran}		
\else		
\documentclass[10pt,final, twocolumn]{IEEEtran}
\fi
\usepackage{times}
\usepackage{amsmath,dsfont}
\usepackage{amssymb,amsthm}
\usepackage{epsfig,verbatim}
\usepackage{subfigure}
\usepackage{setspace}
\usepackage{color}
\usepackage{cite}
\usepackage{epstopdf}
\usepackage{graphics}
\usepackage{accents}
\usepackage{acronym}
\usepackage[bookmarks,colorlinks]{hyperref}
\usepackage{booktabs}
\usepackage{mathtools}
\usepackage{algorithm}
\usepackage{algorithmic}
\usepackage{bbm}
\usepackage{enumitem}

\newtheorem{definition}{Definition}
\newtheorem{theorem}{Theorem}
\newtheorem{corollary}{Corollary}
\newtheorem{proposition}{Proposition}
\newtheorem{lemma}{Lemma}
\newtheorem{remark}{Comment}

\definecolor{NewColor}{rgb}{0,0,0}

\newcommand{\myVec}[1]{{\bf #1}}
\newcommand{\myMat}[1]{\mathbbm{#1}}

\newcommand{\myDetVec}[1]{\myVec{\lowercase{#1}}}
\newcommand{\myRandVec}[1]{\myVec{\uppercase{#1}}}
\newcommand{\myDetMat}[1]{\myMat{\uppercase{#1}}}

\newcommand{\mySet}[1]{\mathcal{#1}}

\newcommand{\E}{\mathcal{E}}		 			
\newcommand{\myU}{{\myRandVec{U}}}			 	
\newcommand{\myW}{{\myRandVec{W}}}			 		
\newcommand{\myY}{{\myRandVec{Y}}}			 		
\newcommand{\myX}{{\myRandVec{X}}}			 		
\newcommand{\myI}{{\myDetMat{i}}}			 		
\newcommand{\myS}{{\myDetMat{s}}}			 		
\newcommand{\myQ}{{\myDetMat{q}}}			 		
\newcommand{\myKron}{\tilde{\myU}}	 		
\newcommand{\Fmat}[1]{\myDetMat{f}_{#1}}	
\newcommand{\tFmat}[1]{\tilde{\myDetMat{f}}_{#1}}	
\newcommand{\CovMat}[1]{\myDetMat{c}_{#1}}			
\newcommand{\myA}{\myDetMat{a}}				
\newcommand{\Avec}{\myDetVec{a}}				
\newcommand{\myP}{P}
\newcommand{\Emat}{\myDetMat{E}}			
\newcommand{\opt}{^{\rm MI}}			
\newcommand{\subopt}{'}		
\newcommand{\nkrp}{^{\rm O}}	
\newcommand{\maskedF}{^{\rm MF}}		
\newcommand{\maskedFU}{^{\rm U}}	
\newcommand{\lenU}{n}			 			
\newcommand{\lenY}{m}			 			
\newcommand{\lenUSet}{\mathcal{N}}			 
\newcommand{\lenYSet}{\mathcal{M}}			 
\newcommand{\SigW}{\sigma_W^2}

\newcommand{\DaMat}{\tilde{\myDetMat{D}}_A}
\newcommand{\Gmat}{\myDetMat{G}}			
\newcommand{\gvec}{\myDetVec{g}}						
\newcommand{\myu}{{\myDetVec{U}}}			 	
\newcommand{\myy}{{\myDetVec{Y}}}			 		

\long\def\symbolfootnote[#1]#2{\begingroup\def\thefootnote{\fnsymbol{footnote}}\footnote[#1]{#2}\endgroup}

\acrodef{dft}[DFT]{discrete Fourier transform}
\acrodef{nb}[NB]{narrowband}
\acrodef{dt}[DT]{discrete-time}
\acrodef{ct}[CT]{continuous-time}
\acrodef{evd}[EVD]{eigenvalue decomposition}
\acrodef{svd}[SVD]{singular value decomposition}
\acrodef{soi}[SOI]{signal of interest}
\acrodef{awgn}[AWGN]{additive white Gaussian noise}
\acrodef{wss}[WSS]{wide-sense stationary}
\acrodef{mse}[MSE]{mean-squared error}
\acrodef{mmse}[MMSE]{minimum MSE}
\acrodef{mi}[MI]{mutual information}
\acrodef{lmmse}[LMMSE]{linear MMSE}
\acrodef{mimo}[MIMO]{multiple input-multiple output}
\acrodef{map}[MAP]{maximum a-posteriori probability}
\acrodef{mi}[MI]{mutual information}
\acrodef{isi}[ISI]{intersymbol interference}
\acrodef{snr}[SNR]{signal-to-noise ratio}
\acrodef{pc}[PC]{proper-complex}
\acrodef{pdf}[PDF]{probability density function}
\acrodef{taf}[TAF]{truncated amplitude flow}
\acrodef{krp}[KRP]{Khatri-Rao product}
\acrodef{rv}[RV]{random variable}
\acrodef{hio}[HIO]{Hybrid Input Output}

 \IEEEoverridecommandlockouts
\title{Measurement Matrix Design for Phase Retrieval Based on Mutual Information 
}

\vspace{-0.5cm}

\author{
	\IEEEauthorblockN{\vspace{-0.2cm} Nir Shlezinger, Ron Dabora, and Yonina C. Eldar\\
	}
	
	\thanks{The work of  R. Dabora is supported in part by the Israel Science Foundation	under Grant no. 1685/16. This paper was presented in part at the 2017 International Symposium on Information Theory.}
 \thanks{N. Shlezinger and Y. C. Eldar are with the Department of Electrical Engineering, Technion –- Israel Institute of Technology, Haifa, Israel (e-mail: nirshlezinge@technion.ac.il ; yonina@ee.technion.ac.il). }	
 \thanks{R. Dabora is with the Department of Electrical and Computer
	 	Engineering, Ben-Gurion University, Beer-Sheva, Israel (e-mail:	 ron@ee.bgu.ac.il).}  	 		
	
	\vspace{-1.0cm}
	
}

\vspace{-0.75cm}

\begin{document}

\maketitle
\begin{abstract}
	In phase retrieval problems, a signal of interest (SOI) is reconstructed based on the magnitude of a linear transformation of the  SOI observed with additive noise. The linear transform is typically referred to as a measurement matrix. 
	Many works on phase retrieval assume that the measurement matrix is a random Gaussian matrix, which, in the noiseless scenario  with sufficiently many measurements, guarantees invertability of the transformation between the SOI and the observations, up to an inherent phase ambiguity.
	However, in many practical applications,  the measurement matrix corresponds to an underlying physical setup, and is therefore deterministic, possibly with structural constraints. 
	In this work we study the design of deterministic measurement matrices, based on maximizing the mutual information between the SOI and the observations. We characterize necessary conditions for the optimality of a measurement matrix, 
	and analytically obtain the optimal matrix in the low signal-to-noise ratio regime. Practical methods for designing general measurement matrices and masked Fourier measurements are proposed. 
	Simulation tests  demonstrate the performance gain achieved by the suggested techniques compared to random Gaussian measurements for various phase recovery algorithms. 
\end{abstract}

\vspace{-0.5cm}
\section{Introduction}
\label{sec:Intro}
\vspace{-0.15cm}
In a wide range of practical scenarios, including X-ray crystallography \cite{Harrison:93}, diffraction imaging \cite{Bunk:07}, astronomical imaging \cite{Dainty:87}, and microscopy \cite{Miao:08}, a \ac{soi} needs to be reconstructed from observations which consist of the magnitudes of its linear transformation with additive noise.  
This class of signal recovery problems is commonly referred to as {\em phase retrieval} \cite{Shechtman:15}.  
%
%
%
In a typical phase retrieval setup, the \ac{soi} is first projected using a {\em measurement matrix} specifically designed for the considered setup. The observations are then obtained as noisy versions of the magnitudes of these projections. Recovery algorithms for phase retrieval received much research attention in recent years. Major approaches for designing phase retrieval algorithms include alternating minimization techniques \cite{Gerchberg:72, Fienup:82}, methods based on convex relaxation, such as {\em phaselift} \cite{Candes:15} and {\em phasecut} \cite{Waldspruger:15}, and non-convex algorithms with a suitable initialization, such as {\em Wirtinger flow} \cite{Candes:15a}, 
and {\em \ac{taf}} \cite{Wang:16}. 

The problem of designing the measurement matrix received considerably less attention compared to the design of phase retrieval algorithms. An important desirable property that measurement matrices should satisfy is a unique relationship between the signal and the magnitudes of its projections, up to an inherent phase ambiguity. 
In many works, particularly in theoretical performance analysis of phase retrieval algorithms \cite{Candes:15, Candes:15a, Candes:14}, the  matrices are assumed to be {\em random}, commonly  with {\rm i.i.d. Gaussian entries}. 
\label{txt:Recovery0}
 However, in practical applications, the measurement matrix corresponds to a fixed physical setup, 
so that it is typically a deterministic matrix, with possibly structural constraints. For example, in optical imaging, lenses are modeled using \ac{dft} matrices and optical masks correspond to diagonal  matrices \cite{Candes:15b}. 
Measurements based on oversampled \ac{dft} matrices were studied in \cite{Huang:16}, measurement matrices which correspond to the parallel application of several \ac{dft}s to modulated versions of the \ac{soi} were proposed  in  \cite{Candes:15},
\textcolor{NewColor}{
and \cite{Dremeau:15}  studied phase recovery using fixed binary measurement matrices, representing hardware limitations in optical imaging systems.} 

All the works above considered {\em noiseless observations}, hence, the focus was on obtaining uniqueness of the magnitudes of the projections in order to guarantee recovery, though the recovery method may be intractable \cite{Eldar:14}. 
When noise is present, such uniqueness no longer guarantees recovery, thus a different design criterion should be considered.
\label{txt:Recovery1}
\textcolor{NewColor}{
Recovery algorithms as well as specialized deterministic measurement matrices were considered in several works. In particular,  
\cite{Jaganathan:16,Bendory:17} studied phase recovery from short-time Fourier transform measurements,
\cite{Pedarsani:15} proposed a recovery algorithm and measurement matrix design based on sparse graph codes for sparse \acp{soi} taking values on a finite set, \cite{Iwen:16} suggested an algorithm using correlation based measurements for flat \acp{soi}, i.e., strictly non-sparse \acp{soi}, and \cite{Bodmann:17} studied recovery methods and the corresponding measurement matrix design for the noisy phase retrieval setup by representing the projections as complex polynomials.}  

A natural optimality condition for the noisy setup, without focusing on a specific recovery algorithm, is to design the measurement matrix to minimize the achievable \ac{mse} in estimating the \ac{soi} from the observations. However, in phase retrieval, the \ac{soi} and observations are not jointly Gaussian, which makes computing the \ac{mmse} for a given measurement matrix in the vector setting very difficult. 
Furthermore, even in the linear non-Gaussian setting, a closed-form expression for the derivative of the \ac{mmse} exists only for the scalar case \cite{Guo:11}, which corresponds to a single observation. Therefore, gradient-based approaches for \ac{mmse} optimization are difficult to apply as well.

In this work we propose an alternative design criterion for the measurement matrix based on maximizing the \ac{mi} between the observations and the \ac{soi}. 
 \ac{mi} is a statistical measure which quantifies the ``amount of information" that one \ac{rv} ``contains" about another \ac{rv} \cite[Ch. 2.3]{Cover:06}. Thus,  maximizing the \ac{mi}  essentially maximizes the statistical dependence between the observations and the \ac{soi}, which is desirable in recovery problems.
\ac{mi} is also related to  \ac{mmse} estimation in Gaussian noise via its derivative \cite{Guo:05}, 
and has been used as the design criterion  in several problems, including the design of projection matrices in compressed sensing \cite{Carson:12} and the construction of radar waveforms \cite{Bell:93, Yang:07}. 

In order to rigorously express the \ac{mi} between the observations and the \ac{soi}, we adopt a Bayesian framework for the phase retrieval setup, similar to the approach in \cite{Riegler:15}. 
Computing  the \ac{mi} between the observations and the \ac{soi} is a  difficult task. Therefore, to facilitate the analysis, we first restate the phase retrieval setup as a linear \ac{mimo} channel of extended dimensions with an additive Gaussian noise. In the resulting \ac{mimo} setup, the channel matrix is given by the row-wise \ac{krp} \cite{Liu:08} of the measurement matrix and its conjugate, while the channel input is the Kronecker product of the \ac{soi} and its conjugate, and is thus non-Gaussian for any \ac{soi} distribution. 
We show that the \ac{mi} between the observations and the \ac{soi} of the original phase retrieval problem is equal to the \ac{mi} between the input and the output of this \ac{mimo} channel. 
Then, we use that fact that  for \ac{mimo} channels with additive Gaussian noise, the gradient of the \ac{mi} can be obtained in closed-form \cite{Palomar:06} {\em for any arbitrary input distribution}. We note that a similar derivation cannot be carried out with the \ac{mmse} design criterion since: 
$1)$ Differently from the \ac{mi}, the \ac{mmse} for the estimation of the \ac{soi} based on the original observations is not equal to the \ac{mmse} for the estimation of the \ac{mimo} channel input based on the output;
$2)$ For the \ac{mimo} setup, a closed-form expression for the gradient of the \ac{mmse} exists only when the input is Gaussian, yet, the input is non-Gaussian for any \ac{soi} distribution due its Kronecker product structure.

Using the equivalent \ac{mimo} channel with non-Gaussian input, we derive necessary conditions on the measurement matrix to maximize the 
 \ac{mi}. 
  We then obtain a closed-form expression for the optimal measurement matrix in the low \ac{snr} regime when the \ac{soi} distribution satisfies a symmetry property, we refer to as Kronecker symmetry, exhibited by, e.g., the zero-mean \ac{pc} Gaussian distribution.
Next, we propose a practical measurement matrix design by approximating the matrix which maximizes the \ac{mi} for any arbitrary \ac{snr}. In our approach, we first maximize the \ac{mi} of a  \ac{mimo} channel, derived from the phase retrieval setup, after relaxing the structure restrictions on the channel matrix imposed by the  phase retrieval problem. We then find the measurement matrix for which the resulting \ac{mimo} channel matrix (i.e., the channel matrix which satisfies the row-wise \ac{krp} structure)  is closest to the \ac{mi} maximizing channel matrix obtained without the structure restriction. 
With this approach, we obtain closed-form expressions for general (i.e., structureless) measurement matrices, as well as for  constrained settings corresponding to masked Fourier matrices, representing, e.g., optical lenses and masks.
The substantial benefits of the proposed design framework are clearly illustrated in a simulations study. 
In particular, we show that our suggested practical design improves the performance of various recovery algorithms compared to using random measurement matrices.

The rest of this paper is organized as follows: 
Section~\ref{sec:Preliminaries} formulates the  problem.
Section~\ref{sec:MiMax}  characterizes necessary conditions on the measurement matrix which maximizes the 
\ac{mi}, and studies its design in the low \ac{snr} regime.
Section~\ref{sec:Algo} presents the proposed approach for designing practical measurement matrices, 
and Section~\ref{sec:Simulations} illustrates the performance of our design in simulation examples.
Finally,  Section~\ref{sec:Conclusions}  concludes the paper.
Proofs of the results stated in the paper are provided in the appendix.

\vspace{-0.2cm}
\section{Problem Formulation}
\label{sec:Preliminaries}
\vspace{-0.15cm}
\subsection{Notations}
\label{subsec:notations}
\vspace{-0.1cm}
We use upper-case letters to denote \ac{rv}s, e.g., $X$, 
lower-case letters for deterministic variables, e.g., $x$,  
and calligraphic letters to denote sets, e.g., $\mathcal{X}$. 
We denote column vectors with boldface letters, e.g., ${\myDetVec{x}}$ for a deterministic vector and ${\myRandVec{X}}$ for a random vector;
the $i$-th element of ${\myDetVec{x}}$  is written as $(\myDetVec{x})_i$. 
Matrices are represented by double-stroke letters,  e.g., $\myDetMat{m}$,  $(\myDetMat{m})_{i,j}$ is the $(i,j)$-th element of $\myDetMat{m}$, and $\myI_{n}$ is the $n \times n$ identity matrix. 
Hermitian transpose, transpose, complex conjugate, real part, imaginary part, stochastic expectation,  and \ac{mi} are denoted by $(\cdot)^H$, $(\cdot)^T$, $(\cdot)^*$, ${\rm Re}\{ \cdot \}$, ${\rm Im}\{ \cdot \}$, $\E\{ \cdot \}$,  and $I\left( \cdot ~ ; \cdot \right)$, respectively. 
${\rm {Tr}}\left(\cdot\right)$ denotes the trace operator, $\left\|\cdot\right\|$ is the Euclidean norm when applied to vectors and the Frobenius norm when applied to matrices, $\otimes$ denotes the Kronecker product, 
$\delta_{k,l}$ is the Kronecker delta function, i.e., $\delta_{k,l}\! =\! 1$ when $k\!=\!l$ and $\delta_{k,l}\! =\! 0$ otherwise, 
and $a^+ \!\triangleq\! \max\{0,a\}$. 
For an $n\times 1$ vector $\myDetVec{x}$, ${\rm diag}\left( \myDetVec{x}\right)$ is the $n\times n$ diagonal matrix whose diagonal entries are the elements of $\myDetVec{x}$, i.e., $\left( {\rm diag}\left( \myDetVec{x}\right)\right)_{i,i}= \left( \myDetVec{x}\right)_i$.
The sets of real and of complex numbers are denoted by $\mySet{R}$  and $\mySet{C}$, respectively. 
Finally, for an $n \! \times \! n$ matrix $\myDetMat{x}$,  $\myDetVec{x}\! =\! {\rm vec}\left(\myDetMat{x}\right)$ is the $n^2 \! \times \! 1$ column vector obtained by stacking the columns of $\myDetMat{x}$ one below the other. 
 The  $n \! \times \! n$ matrix  $\myDetMat{x}$ is recovered from $\myDetVec{x}$ via $\myDetMat{x} = {\rm vec}_n^{-1} \!\left(\myDetVec{x}\right)$. 

%
\vspace{-0.2cm}
\subsection{The Phase Retrieval Setup}
\label{subsec:Pre_Model}
\vspace{-0.1cm}
We consider the recovery of a random \ac{soi} $\myU \in \mySet{C}^{\lenU}$, from an observation vector  $\myY \in \mySet{R}^{\lenY}$. Let $\myA \in \mySet{C}^{\lenY \times \lenU}$ be the measurement matrix and $\myW \in \mySet{R}^{\lenY}$ be the additive noise, modeled as a zero-mean real-valued Gaussian vector with covariance matrix $\SigW \myI_{\lenY}$,  $\SigW > 0$. 
As in  \cite[Eq. (1.5)]{Candes:14}, \cite[Eq. (1)]{Huang:16}, and \cite[Eq. (1.1)]{Eldar:14}, the relationship between $\myU$ and $\myY$  is given by:
\vspace{-0.1cm}
\begin{equation}
\label{eqn:model1}
\myY = \left|\myA \myU \right|^2 + \myW,
\vspace{-0.1cm}
\end{equation}
where $\left|\myA \myU \right|^2$ denotes {\em the element-wise squared magnitude}. 
Since for every $\theta \in \mySet{R}$, the vectors $\myU$ and $\myU e^{j \theta}$ result in the same $\myY$, the vector $\myU$ can be recovered only up to a global phase. 
 
In this work we study the design of $\myA$ aimed at maximizing  the \ac{mi} between the \ac{soi} and the observations. 
Letting $f(\myu, \myy)$ be the joint \ac{pdf} of $\myU$ and $\myY$, $f(\myu)$  the \ac{pdf} of $\myU$, and $f(\myy)$  the \ac{pdf} of $\myY$, the \ac{mi} between the \ac{soi} $\myU$ and the observations $\myY$ is given by \cite[Ch. 8.5]{Cover:06} 
\begin{equation}
\label{eqn:MIDef}
I\left(\myU;\myY\right) \triangleq \E_{\myU, \myY}\left\{\log\frac{f(\myU, \myY)}{f(\myU)f( \myY)} \right\}.
\end{equation}
 Specifically, we  study the measurement matrix $\myA\opt$ which maximizes\footnote{\label{ftn:NonUnique}The optimal matrix  $\myA\opt$ is not unique since, for example, for any real $\phi$, the matrices $\myA$ and  $\myA e^{j\phi}$ result in the same \ac{mi} $I\left(\myU;\myY\right)$.} the \ac{mi}  for a fixed arbitrary distribution of $\myU$, subject to a Frobenious norm constraint $\myP > 0$, namely, 
\vspace{-0.1cm}
\begin{equation}
\label{eqn:MIProbDef1}
\myA\opt= \mathop {{\rm{argmax}}}\limits_{\myA \in \mySet{C}^{\lenY \times \lenU}:{\rm{ }}{\rm{Tr}}\left( \myA\myA^H \right) \le \myP} I\left(\myU;\myY\right),
\vspace{-0.1cm}
\end{equation}
where $\myU$ and $\myY$ are related via \eqref{eqn:model1}. 
In the noiseless non-Bayesian phase retrieval setup, it has been shown that a necessary and sufficient condition for the existence of a bijective mapping from $\myU$ to $\myY$ is that the number of observations, $\lenY$, is linearly related to the dimensions of the \ac{soi}\footnote{Specifically, $\lenY = 4\lenU - 4$ was shown to be sufficient and $\lenY = 4\lenU-\mathcal{O}(\lenU)$ was shown to be necessary.}, $\lenU$, see \cite{Heinosaari:13, Bodmann:15}. Therefore, we focus on values of $\lenY$ satisfying  $\lenU\! \leq\! \lenY\! \leq\! \lenU^2$.

As discussed in the introduction, in practical scenarios, the structure of the measurement matrix is often constrained. One type of structural constraint commonly encountered in practice is the masked Fourier structure, which arises, for example, when the measurement matrix represents an optical setup consisting of lenses and masks \cite{Candes:15b,Pedarsani:15}.
In this case, $\myY$ is obtained by projecting $\myU$ via $b$ optical masks,  each modeled as an $\lenU \times \lenU$ diagonal matrix $\Gmat_{l}$, $l \in \{1,2,\ldots,b\} \triangleq \mathcal{B}$, followed by an optical lens, modeled as a \ac{dft} matrix of size $\lenU$, denoted $\Fmat{\lenU}$ \cite[Sec. 3]{Pedarsani:15}. Consequently, $\lenY \!= \!b \cdot \lenU$ and $\myA$ is obtained as
\vspace{-0.1cm}
\begin{equation}
\label{eqn:KRP2AMatForm}
\myA 
= \left[ {\begin{array}{*{20}{c}}
	\Fmat{\lenU}\Gmat_{1} \\
	\Fmat{\lenU}\Gmat_{2} \\
	\vdots \\
	\Fmat{\lenU}\Gmat_{b} 
	\end{array}} \right] 
= \left(\myI_b \otimes	\Fmat{\lenU}  \right) \left[ {\begin{array}{*{20}{c}}
	\Gmat_{1}\\
	\Gmat_{2}\\
	\vdots \\
	\Gmat_{b}
	\end{array}} \right].
\vspace{-0.1cm}
\end{equation}
Since $ \lenU \le \lenY \le \lenU^2$, we focus on $1 \le b \le \lenU$. 
In the following sections we study the optimal design of general (unconstrained) measurement matrices, and propose a practical algorithm for designing both general measurement matrices as well as masked Fourier measurement matrices.  

\vspace{-0.25cm}
\section{Optimal Measurement Matrix}
\label{sec:MiMax}
\vspace{-0.1cm}
In this section we first  show that the relationship \eqref{eqn:model1} can be equivalently represented (in the sense of having the same \ac{mi}) as a \ac{mimo} channel with \ac{pc} Gaussian noise. Then, we use the equivalent representation to study the design of measurement matrices for two cases: The first  considers an arbitrary \ac{soi} distribution, for which we characterize a necessary condition on the optimal measurement matrix. The second case treats an \ac{soi} distribution satisfying a symmetry property (exhibited by, e.g., zero-mean \ac{pc} Gaussian distributions) focusing on the low \ac{snr} regime, for which we obtain the optimal measurement matrix in closed-form.

\vspace{-0.35cm}
\subsection{Gaussian \ac{mimo} Channel Interpretation}
\label{subsec:MIMORep}
\vspace{-0.1cm}
In order to characterize the solution of \eqref{eqn:MIProbDef1}, we first consider the relationship \eqref{eqn:model1}: Note that for every $p \in \{1,2,\ldots,\lenY\} \triangleq \lenYSet$, the $p$-th entry of $ \left|\myA \myU \right|^2$ can be written as
\vspace{-0.1cm}
\begin{equation}
\label{eqn:AUdef1}
\left( \left|\myA \myU \right|^2\right)_p    
= \sum\limits_{k = 1}^\lenU {\sum\limits_{l = 1}^\lenU {\left( \myA  \right)}_{p,k}\left( \myA  \right)_{p,l}^* {{\left( \myU \right)}_k}\left( \myU\right)_l^*}.
\vspace{-0.1cm}
\end{equation}
Next, define $\lenUSet \triangleq \{1,2,\ldots, \lenU\}$, and the $\lenY \times \lenU^2$ matrix $\tilde{\myA}$ such that 
\vspace{-0.1cm}
\begin{equation}
\label{eqn:AtildeDef}
\big( \tilde{\myA}\big)_{p,(k-1) \lenU + l} \triangleq   {\left( \myA  \right)}_{p,k}\left( \myA  \right)_{p,l}^*, \quad p \in \lenYSet,\;\; k,l \in \lenUSet.
\vspace{-0.1cm}
\end{equation}
Letting  $\myKron \!\triangleq\! \myU \otimes \myU^*$,  from \eqref{eqn:AUdef1} we obtain that $\left|\myA \myU \right|^2  \!=\! \tilde{\myA}\left(\myU \!\otimes \!\myU^* \right)$. Thus \eqref{eqn:MIProbDef1} can be written as
\vspace{-0.1cm}
\begin{equation}
\label{eqn:AUdef2}
\myY =  \tilde{\myA}\left(\myU \otimes \myU^* \right) + \myW \equiv \tilde{\myA}\myKron + \myW.
\vspace{-0.1cm}
\end{equation}
We note that the transformation from $\myU$ to $\myKron = \myU \otimes \myU^*$ is bijective\footnote{The transformation from $\myU$ to $\myKron$ is bijective up to a global phase. However, the global phase can be set to an arbitrary value, as \eqref{eqn:model1} is not affected by this global phase. Therefore, bijection up to a global phase is sufficient for establishing equivalence of the two representations in the present setup.},
since $\myU$ can be obtained from the \ac{svd} of the rank one matrix  $\myU \myU^H = {\rm vec}_\lenU^{-1}\!\left( \myU \otimes \myU^*\right)^T$  \cite[Ch. 2.4]{Golub:13}.
We also note that $\tilde{\myA}$ corresponds to the {\em row-wise \ac{krp}} of $\myA$ and $\myA^*$ \cite[Ch. 12.3]{Golub:13}, namely, the rows of $\tilde{\myA}$ are obtained as the Kronecker product of the corresponding rows of  $\myA$ and $\myA^*$. Defining $\myS_{\lenY}$ to be the $\lenY \times \lenY^2$ selection matrix such that $\left( \myS_{\lenY}\right)_{k,l} \!=\! \delta_{l,(k\!-\!1)\lenY\! +\! k}$, we can write $\tilde{\myA}$ as \cite[Sec. 2.2]{Liu:08} 
\vspace{-0.1cm}
\begin{equation}
\label{eqn:AtildeDef2}
\tilde{\myA} = \myS_{\lenY} \cdot \left( \myA \otimes \myA^* \right).
\vspace{-0.1cm}
\end{equation}
%

The relationship \eqref{eqn:AUdef2} formulates the phase retrieval setup as {\em  a \ac{mimo} channel} with complex channel input $\myKron$, complex channel matrix $\tilde{\myA}$, real additive Gaussian noise $\myW$, and real channel output $\myY$.  
We note that   $\myKron = \myU \otimes \myU^*$ is non-Gaussian for any distribution of $\myU$, since, e.g., $ \big( \myKron\big)_1 = \left| \left( \myU \right)_1\right|^2$ is non-negative.
In order to identify the measurement matrix which maximizes the \ac{mi}, we wish to apply the gradient of the \ac{mi} with respect to the measurement matrix, stated in \cite[Thm. 1]{Palomar:06}. 
To facilitate this application, we next formulate the phase retrieval setup as {\em a complex \ac{mimo} channel} with {\em additive \ac{pc} Gaussian noise}. To that aim, let $\myW_I \in \mySet{R}^\lenY$ be a random vector, distributed identically to $\myW$ and independent of both $\myW$ and $\myU$, and also let $\myY_C \triangleq \myY + j\myW_I$. The relationship between $\myY_C$ and $\myKron$ corresponds to a complex \ac{mimo} channel with additive zero-mean \ac{pc} Gaussian noise, $\myW_C \triangleq \myW + j\myW_I$, with covariance matrix $2\SigW \myI_{\lenY}$:  
\vspace{-0.1cm}
\begin{equation}
\myY_C = \tilde{\myA}\myKron + \myW_C.
\label{eqn:MIEq0}
\vspace{-0.1cm}
\end{equation}
As the mapping from $\myU$ to $\myKron$ is bijective, it follows from \cite[Corollary after Eq. (2.121)]{Cover:06} that 
\vspace{-0.1cm}
\begin{equation}
I\big(\myU;\myY\big) 
= I\big(\myKron;\myY\big) 
\stackrel{(a)}{=} I\big(\myKron;\myY_C\big),
\label{eqn:MIEq1}
\vspace{-0.1cm}
\end{equation}
where $(a)$ follows from the \ac{mi} chain rule \cite[Sec. 2.5]{Cover:06}, since $\myY \!=\! {\rm Re}\left\{\myY_C \right\}$, $\myW_I\! =\! {\rm Im}\left\{\myY_C \right\}$, and $\myW_I$ is independent of $\myY$ and $\myU$.   Thus, \eqref{eqn:MIProbDef1} can be solved by finding $\myA$ which maximizes the input-output \ac{mi} of the \ac{mimo} channel representation. 

The \ac{mimo} channel interpretation represents the non-linear phase retrieval setup \eqref{eqn:model1} as a linear problem \eqref{eqn:MIEq0} {\em without modifying the \ac{mi}}. This presents an advantage of using \ac{mi} as a design criterion over the \ac{mmse}, as, unlike \ac{mi},  \ac{mmse} is not invariant to the linear representation, i.e., the error covariance matrices of the \ac{mmse} estimator of $\myU$ from $\myY$ and of the \ac{mmse} estimator of $\myKron$ from $\myY_C$ are in general not the same.

\iflowsnr
\vspace{-0.2cm}
\subsection{Conditions on $\myA\opt$ for Arbitrary \ac{soi} Distribution}
\label{subsec:GenInput}
\vspace{-0.1cm}
\fi
	Let $\Emat\left(\myA \right)$ be the error covariance matrix of the \ac{mmse} estimator of $\myKron$ from $\myY$ (referred to henceforth as the {\em \ac{mmse} matrix}) for a fixed measurement matrix 
	 $\myA$, i.e.,
	 \vspace{-0.1cm}
	\begin{equation}
	\Emat\left(\myA \right)  \triangleq \E \Big\{\left(\myKron - \E\big\{\myKron | \myY \big\} \right) \left(\myKron - \E\big\{\myKron | \myY \big\} \right)^H \Big\}.
	\label{eqn:EmatDef}
	\end{equation}
 Based on the observation that \eqref{eqn:MIEq0} corresponds to a \ac{mimo} channel with additive Gaussian noise, we obtain the following necessary condition on $\myA\opt$ which solves \eqref{eqn:MIProbDef1}:
\begin{theorem}[Necessary condition]
	\label{thm:GeneralSource}
	Let
	$\Avec\opt_k$ be the $k$-th column of $\left( \myA\opt\right)^T $, $k \!\in\!  \lenYSet$, and define the $\lenU \times \lenU$ matrix
	\vspace{-0.2cm}
	\begin{align*}
	\!\!\myDetMat{H}_k\!\left(\myA\opt \right)  \!
	&\!\triangleq\! \left( \myI_n\! \otimes \!\left( \Avec\opt_{k}\right)^T\right) \Big( \Emat\!\left(\myA\opt \right)\Big)^T  \left(\myI_n \! \otimes \left( \Avec\opt_{k}\right)^* \right) \notag \\
	&\qquad \!+\! \left( \left( \Avec\opt_{k}\right)^T\! \otimes\! \myI_n \right) \Emat\left(\myA\opt \right) \left( \left( \Avec\opt_{k}\right)^*\! \otimes\! \myI_n  \right)\!.
	\vspace{-0.1cm}
	\end{align*}
	Then,   $\myA\opt$ that solves \eqref{eqn:MIProbDef1} satisfies:
	\vspace{-0.1cm}
	\begin{equation}
	\label{eqn:GeneralSource}
\lambda\Avec\opt_{k}  = \myDetMat{H}_k\!\left(\myA\opt \right) \Avec\opt_{k}, \qquad \forall k \!\in\!  \lenYSet,
	\vspace{-0.1cm}
	\end{equation}
	 where $\lambda \ge 0$ is selected such that ${\rm{Tr}}\!\left(\! \myA\opt\left( \myA\opt\right)^H \right)\! =\! \myP$.   
\end{theorem}
%
{\em Proof:}
	See Appendix \ref{app:GeneralSource}.

\smallskip
It follows from \eqref{eqn:GeneralSource} that the  $k$-th row of $\myA\opt$, $k \in  \lenYSet$, is an eigenvector of the $\lenU \times \lenU$ Hermitian positive semi-definite matrix $\myDetMat{H}_k\!\left(\myA\opt \right)$, which depends on $\myA\opt$. 
As the optimization problem in \eqref{eqn:MIProbDef1} is generally non-concave, condition \eqref{eqn:GeneralSource} does not uniquely identify the optimal measurement matrix in general.
Furthermore, in order to explicitly obtain  $\myA\opt$ from \eqref{eqn:GeneralSource}, the  \ac{mmse} matrix $\Emat\left(\myA\opt \right)$ must be derived, which is not a simple task. As an example, let the entries of  $\myU$ be zero-mean i.i.d. \ac{pc} Gaussian \ac{rv}s. Then, $\myKron$ obeys a singular Wishart distribution \cite{Ratnarajah:05}, and $\Emat\left(\myA \right)$ does not seem to have a tractable analytic expression. 
\iflowsnr
Despite this general situation, when the \ac{snr} is sufficiently low, we can explicitly characterize $\myA\opt $ in certain scenarios, as discussed in the next subsection.

\vspace{-0.35cm}
\subsection{Low SNR Regime}
\label{subsec:LowSNR}
\vspace{-0.1cm}
We next show that in the low \ac{snr} regime, it is possible to obtain an expression for the optimal measurement matrix which does not depend on  $\Emat\left( \myA \right)$.
\label{txt:Mod1}
\textcolor{NewColor}{
Let $\CovMat{\myU}$ and $\CovMat{\myKron}$ denote the covariance matrices of the \ac{soi}, $\myU$, and of $\myKron = \myU \otimes \myU^*$, respectively.} In the low \ac{snr} regime, i.e., when $\frac{\myP}{\SigW} \rightarrow 0$, 
the \ac{mi} $I\big(\myKron;\myY_C\big)$ satisfies  \cite[Eq. (41)]{Palomar:06}: 
\vspace{-0.1cm}
\begin{equation}
\label{eqn:MIProbDef2a}
I\left(\myKron;\myY_C\right) \approx \frac{1}{2\SigW}{\rm Tr}\left(\tilde{\myA}\CovMat{\myKron} \tilde{\myA}^H \right). 
\vspace{-0.1cm}
\end{equation}
Thus, from \eqref{eqn:MIEq1} and \eqref{eqn:MIProbDef2a}, the measurement matrix maximizing the \ac{mi} in the low \ac{snr} regime can be approximated by 
\vspace{-0.1cm}
\begin{equation}
\myA\opt 
\approx  \mathop {{\rm{argmax}}}\limits_{\myA\in \mySet{C}^{\lenY \times \lenU}:{\rm{ }}{\rm{Tr}}\left( \myA\myA^H \right) \le \myP} {\rm Tr}\left(\tilde{\myA}\CovMat{\myKron} \tilde{\myA}^H \right),
\label{eqn:MIProbDef2}
\vspace{-0.1cm}
\end{equation}
\label{txt:Mod1a}
\textcolor{NewColor}{
where $\tilde{\myA}$ is given by \eqref{eqn:AtildeDef2}.}

Next, we introduce a new concept we refer to as Kronecker symmetric random vectors:
\begin{definition}[Kronecker symmetry]
	\label{def:KronSym}
	A random vector $\myX$ with covariance matrix $\CovMat{\myX}$ is said to be {\em Kronecker symmetric} if the covariance matrix of $\myX \otimes \myX^*$ is equal to $\CovMat{\myX} \otimes \CovMat{\myX}^*$.
\end{definition}
In particular, zero-mean \ac{pc} Gaussian distributions satisfy Def. \ref{def:KronSym}, as stated in the following lemma:
\begin{lemma}
	\label{lem:Cond2}
	Any $\lenU \times 1$ zero-mean \ac{pc} Gaussian random vector is Kronecker symmetric.
%
\end{lemma}
%
	{\em Proof:}
	See Appendix \ref{app:Cond2}.
%

\smallskip
We now obtain a closed-form solution to  \eqref{eqn:MIProbDef2}  when $\myU$ is a Kronecker symmetric random vector. The optimal $\myA\opt$ for this setup is stated in the following theorem:
\begin{theorem}
	\label{thm:GaussianSource}
		Let 	$\Avec\opt_k$ be the $k$-th column of $\left( \myA\opt\right)^T $, $k \!\in\!  \lenYSet$, and let $\myVec{v}_{\max}$ be the eigenvector of $\CovMat{\myU}$ corresponding to its maximal eigenvalue. 
	If $\myU$ is a Kronecker symmetric random vector with covariance matrix $\CovMat{\myU}$, then, for every $\myVec{c} \! \in\! \mySet{C}^{\lenY}$ with $\|\myVec{c}\|^2\! =\! \myP$, setting $\Avec\opt_k\! =\! \left( \myVec{c} \right)_k  \myVec{v}_{\max}^*$ for all  $k \!\in\!  \lenYSet$ solves \eqref{eqn:MIProbDef2}. Thus, 
	\vspace{-0.1cm}
	\begin{equation}
	\label{eqn:GaussianSource}
	\myA\opt = \myVec{c}\cdot\myVec{v}_{\max}^H. 
	\vspace{-0.1cm}
	\end{equation}
%
%
\end{theorem} 
%
	{\em Proof:}
	See Appendix \ref{app:GaussianSource}.
%

\smallskip
The result of Theorem \ref{thm:GaussianSource} is quite non-intuitive from an estimation perspective, as it suggests using {\em a rank-one measurement matrix}. This implies that the optimal measurement matrix projects the multivariate \ac{soi} onto a single eigenvector corresponding to the largest spread. Consequently, there are infinitely many realizations of $\myU$ which result in the same $\left|\myA\myU\right|^2$.  
The optimality of rank-one measurements can be explained by noting that the selected scalar projection is, in fact, the least noisy of all possible scalar projections, as it corresponds to the largest eigenvalue of the covariance matrix of the \ac{soi}. Hence, when the additive noise is dominant, the optimal strategy is to design the measurement matrix such that it keeps only the least noisy spatial dimension of the signal, and eliminates all other spatial dimensions which are very noisy. 
From an information theoretic perspective, this concept is not new, and the strategy of using a single spatial dimension which corresponds to the largest eigenvalue of the channel matrix in memoryless \ac{mimo} channels was shown to be optimal in the low \ac{snr} regime, e.g., in the design of the optimal precoding matrix for \ac{mimo} Gaussian channels \cite[Sec II-B]{Cruz:10}. However, while in \cite[Sec II-B]{Cruz:10} the problem was to optimize the input covariance (using the precoding matrix) for a {\em given channel}, in our case we optimize over the ``channel" (represented by the measurement matrix) for a {\em given \ac{soi} covariance matrix}.

Finally, we show that the optimal measurement matrix in  Theorem \ref{thm:GaussianSource} satisfies the necessary condition for optimality in Theorem \ref{thm:GeneralSource}: In the low \ac{snr} regime the \ac{mmse} matrix \eqref{eqn:EmatDef} satisfies $\Emat\big(\myA\big) \approx \CovMat{\myKron}$, see, e.g., \cite[Eq. (15)]{Cruz:10}. The Kronecker symmetry of the \ac{soi} implies that $\Emat\big(\myA\big) \approx \CovMat{\myU}\otimes \CovMat{\myU}^*$. Plugging this into the definition of $\myDetMat{H}_k\!\left(\myA\opt \right)$ in Theorem \ref{thm:GeneralSource} results in $\myDetMat{H}_k\!\left(\myA\opt \right) = 2 \left( \left( \Avec\opt_{k}\right)^T \CovMat{\myU}\left( \Avec\opt_{k}\right)^* \right)\CovMat{\myU}^*$. Theorem \ref{thm:GeneralSource} thus states that  for every $k \in \lenYSet$, the vector $\Avec\opt_{k}$ must be a complex conjugate of an eigenvector of $\CovMat{\myU}$. Consequently, the optimal matrix in Theorem \ref{thm:GaussianSource} satisfies the necessary condition in Theorem \ref{thm:GeneralSource}.
\fi

\vspace{-0.25cm}
\section{Practical Design of the Measurement Matrix}
\label{sec:Algo}
\vspace{-0.1cm}
As can be concluded from the discussion following Theorem~\ref{thm:GeneralSource}, the fact that \eqref{eqn:GeneralSource} does not generally have a unique solution combined with the fact that it is often difficult to analytically compute the \ac{mmse} matrix,  make the characterization of the optimal measurement matrix from condition \eqref{eqn:GeneralSource} a very difficult task.   
Therefore, in this section we propose a practical approach for designing measurement matrices based on Theorem~\ref{thm:GeneralSource}, while circumventing the difficulties discussed above by applying appropriate approximations. 
\label{txt:Poissson}
\textcolor{NewColor}{
We note that while the practical design approach proposed in this section assumes that the observations are corrupted by an additive Gaussian noise, the suggested approach can also be used as an ad hoc method for designing measurement matrices for phase retrieval setups with non-Gaussian noise, e.g., Poisson noise \cite[Sec. 2.3]{Candes:15}.}
The practical design is performed via the following steps:
First, 
we find the matrix $\tilde{\myA}\opt$ which maximizes the \ac{mi} {\em without restricting $\tilde{\myA}$ to satisfy the row-wise \ac{krp} structure \eqref{eqn:AtildeDef2}}. Ignoring the structural constraints on $\tilde{\myA}$  facilitates characterizing $\tilde{\myA}\opt$ via a set of fixed point equations. 
Then, we obtain a closed-form approximation of $\tilde{\myA}\opt$ by using the  covariance matrix of the \ac{lmmse} estimator instead of the actual \ac{mmse} matrix. We denote the resulting matrix by $\tilde{\myA}\subopt$.
Next, noting that the \ac{mi} is invariant to unitary transformations, we obtain the final  measurement matrix by finding $\myA$ which minimizes the Frobenious norm between $\myS_{\lenY}\big(\myA\! \otimes\! \left( \myA\right) ^*\! \big)$ and a given unitary transformation of $\tilde{\myA}\subopt$, also designed to minimize the Frobenious norm. Using this procedure we obtain closed-form expressions for general measurement matrices as well as for masked Fourier measurement matrices.
In the following we elaborate on these steps.


\vspace{-0.2cm}
\subsection{Optimizing  without Structure Constraints}
\label{subsec:AlgoPart1}
\vspace{-0.1cm}
\label{txt:Mod2}
\textcolor{NewColor}{
In the first step we replace the maximization of the \ac{mi} in \eqref{eqn:MIProbDef1} with respect to the measurement matrix $\myA$, with a maximization with respect to $\tilde{\myA}$, which denotes the row-wise \ac{krp} of $\myA$ and $\myA^*$. Specifically, we look for the matrix  $\tilde{\myA}$ which maximizes  $ I\left(\myKron;\myY_C\right)$,  without constraining the structure of $\tilde{\myA}$, while satisfying the trace constraint in \eqref{eqn:MIProbDef1}.} 

We now formulate a constraint on $\tilde{\myA}$ which guarantees that the trace constraint in \eqref{eqn:MIProbDef1} is satisfied. 
 Letting $\Avec_k$ be the $k$-th column of $\myA^T $, $k \!\in\!  \lenYSet$, we have that
 \vspace{-0.1cm}
 \begin{align}
 \hspace{-0.2cm}
 \big\|{\myA}\big\|^4 
 &=  \sum\limits_{k_1\!=\!1}^\lenY \sum\limits_{k_2\!=\!1}^\lenY  \|\Avec_{k_1}\|^2  \|\Avec_{k_2}\|^2 \notag \\
 &\stackrel{(a)}{\le} \frac{1}{2}\sum\limits_{k_1\!=\!1}^\lenY \sum\limits_{k_2\!=\!1}^\lenY \!\Big( \|\Avec_{k_1}\|^4 \! +\! \|\Avec_{k_2}\|^4  \Big)
 \!=\! \lenY \sum\limits_{k\!=\!1}^\lenY  \|\Avec_{k}\|^4,
 \label{eqn:normineq1}
 \end{align}
 where $(a)$ follows since $a^2 + b^2 \ge 2ab$ for all $a,b \in \mySet{R}$.
 Next, it follows from \eqref{eqn:AtildeDef2} that  
 \vspace{-0.1cm}
 \begin{align}
 \big\|\tilde{\myA}\big\|^2 
 &= \sum\limits_{k=1}^\lenY \|\Avec_k \otimes \Avec_k^*\|^2 \notag \\
 &\stackrel{(a)}{=}    \sum\limits_{k=1}^\lenY \|\Avec_k\|^4  \stackrel{(b)}{\ge} \frac{1}{\lenY}  \|{\myA}\|^4,
 \label{eqn:normineq2}
 \end{align}
 where $(a)$ follows from \cite[Pg. 709]{Golub:13} and $(b)$ follows from \eqref{eqn:normineq1}. 
Therefore, if  $\tilde{\myA}$ satisfies $\big\|\tilde{\myA}\big\|  \le \frac{\myP}{\sqrt{\lenY}}$, then  ${\rm Tr}\left( \myA\myA^H\right) =\|{\myA}\|^2 \leq \myP$, thereby satisfying the constraint in  \eqref{eqn:MIProbDef1}. 
Consequently, we consider the following optimization problem:
\vspace{-0.1cm}
	\begin{equation}
	\tilde{\myA}\opt 
	= \mathop {{\rm{argmax}}}\limits_{\tilde{\myA}\in \mySet{C}^{\lenY \times \lenU^2}:{\rm{ }}{\rm{Tr}}\left( \tilde{\myA}\tilde{\myA}^H \right) \le \frac{\myP^2}{\lenY}} I\left(\myKron;\myY_C\right). 
		\label{eqn:MIProbDef1a}
		\vspace{-0.2cm}
	\end{equation}

\label{txt:Mod3}
\textcolor{NewColor}{
	Note that without constraining   $\tilde{\myA}$ to satisfy the structure \eqref{eqn:AtildeDef2}, $\myY$ can be complex,  and the \ac{mi} between the input and the output of the transformed \ac{mimo} channel, $I\big(\myKron;\myY_C\big)$, may not be equal to the \ac{mi} between the \ac{soi} and the observations of the original phase retrieval setup, $I\left(\myU;\myY\right)$.} 
	%
%
%
%

	The solution to \eqref{eqn:MIProbDef1a} is given in the following lemma:
	\begin{lemma}
		\label{lem:optTildeA1}
	{\bf \cite[Thm. 4.2]{Carson:12}, \cite[Thm. 1]{Lamarca:09},  \cite[Prop. 2]{Payaro:09}:} 
	Let  $\Emat\big(\tilde{\myA} \big)$ be the covariance matrix of the  \ac{mmse} estimate of  $\myKron$ from $\myY_C$ for a given $\tilde{\myA}$, and let $\myDetMat{V}_E\big(\tilde{\myA} \big)  \myDetMat{D}_E\big(\tilde{\myA} \big) \Big(\myDetMat{V}_E\big(\tilde{\myA} \big)\Big)^H$ be the eigenvalue decomposition of $\Emat\big(\tilde{\myA} \big)$, in which  $\myDetMat{V}_E\big(\tilde{\myA} \big)$ is unitary and  $\myDetMat{D}_E\big(\tilde{\myA} \big)$ is a diagonal matrix whose diagonal entries are the eigenvalues of $\Emat\big(\tilde{\myA} \big)$  in descending order.   
	 Let $\myDetMat{D}_A\big(\tilde{\myA} \big)$ be an $\lenY \times \lenU^2$  diagonal matrix whose entries satisfy 
	 \begin{subequations}
	 	\label{eqn:DmatCond}
	 \begin{align}	
	 \vspace{-0.12cm}
	 &\left( \myDetMat{D}_A\big(\tilde{\myA} \big)\right)_{k,k} = 0 \quad {\rm if} \quad \Big( \myDetMat{D}_E\big(\tilde{\myA} \big)\Big) _{k,k} < \eta \\
	 &\left( \myDetMat{D}_A\big(\tilde{\myA} \big)\right)_{k,k} > 0 \quad {\rm if} \quad  \Big( \myDetMat{D}_E\big(\tilde{\myA} \big)\Big) _{k,k} = \eta,
	 \vspace{-0.2cm}
	 \end{align}
	 \end{subequations}
	 where $\eta$ is selected such that $\sum\limits_{k=1}^{\lenY} \Big( \myDetMat{D}_A\big(\tilde{\myA} \big)\Big)_{k,k}^2 = \frac{\myP^2}{\lenY}$. 
	 The matrix $\tilde{\myA}\opt$ which solves \eqref{eqn:MIProbDef1a} is given by the solution to
	 \vspace{-0.1cm}
	 \begin{equation}
	 \label{eqn:optTildeA1}
	 \tilde{\myA}\opt = \myDetMat{D}_A\big(\tilde{\myA}\opt  \big)  \Big(\myDetMat{V}_E\big(\tilde{\myA}\opt  \big)\Big)^H.
	 \vspace{-0.1cm}
	 \end{equation}
	\end{lemma}

Lemma \ref{lem:optTildeA1} characterizes $\tilde{\myA}\opt$ via a set of fixed point 
equations\footnote{The solution in \cite[Thm. 4.2]{Carson:12} includes a permutation matrix which performs mode alignment. However, for white noise mode alignment is not needed, and the permutation matrix  can be set to $\myI_{\lenU^2}$ \cite[Sec. III]{Lamarca:09}.}. 
Note that the  matrix $\myDetMat{D}_A( \tilde{\myA}\opt)$ is constructed such that $\tilde{\myA}\opt$ which solves \eqref{eqn:optTildeA1}  induces a covariance matrix of the \ac{mmse} estimate
 of  $\myKron$ from $\myY_C$, denoted ${\myDetMat{E}}( \tilde{\myA}\opt)$, 
 whose eigenvalues satisfy \eqref{eqn:DmatCond}.

\vspace{-0.2cm}
\subsection{Replacing the \ac{mmse} Matrix with the \ac{lmmse} Matrix}
\label{subsec:AlgoPart2}
\vspace{-0.1cm}
In order to obtain $\tilde{\myA}\opt$ from Lemma \ref{lem:optTildeA1}, we need the error covariance matrix of the \ac{mmse} estimator of $\myKron$ from $\myY_C$, ${\myDetMat{E}}\big(\tilde{\myA}\opt\big)$,  which in turn depends on $\tilde{\myA}\opt$. As ${\myDetMat{E}}\big(\tilde{\myA}\big)$ is difficult to compute, we propose to replace the error covariance matrix of the \ac{mmse} estimate with that of the  \ac{lmmse} estimate\footnote{An inspiration for this approximation stems from the fact that for parallel Gaussian \ac{mimo} scenarios, the covariance matrices of the \ac{mmse} estimate and of the \ac{lmmse} estimate coincide at high \ac{snr}s \cite{Bustin:13}.} of $\myKron$ from $\myY_C$. 
\iflowsnr
\else
Let $\CovMat{\myKron}$ be the covariance matrix of $\myKron$. 
\fi 
The  \ac{lmmse} matrix is given by  \cite[Sec. IV-C]{Palomar:06} 
	\vspace{-0.1cm}
	\begin{equation*}
	{\myDetMat{E}}_L\left(\tilde{\myA} \right) = \CovMat{\myKron} - \CovMat{\myKron}\tilde{\myA}^H \left(2\SigW\myI_{\lenY} + \tilde{\myA}\CovMat{\myKron}\tilde{\myA}^H \right)^{-1} \tilde{\myA}\CovMat{\myKron}.
	\vspace{-0.1cm}
	\end{equation*} 
Replacing  ${\myDetMat{E}}\big(\tilde{\myA} \big)$ with ${\myDetMat{E}}_L\big(\tilde{\myA} \big)$ in Lemma~\ref{lem:optTildeA1}, we obtain the  matrix $\tilde{\myA}\subopt$ stated in the following corollary:
\begin{corollary}
	\label{pro:waterfil1}
	Let $\myDetMat{V}_{\myKron}  \myDetMat{D}_{\myKron}  \myDetMat{V}_{\myKron}^H $ be the eigenvalue decomposition of $\CovMat{\myKron}$, in which $\myDetMat{V}_{\myKron}$ is unitary and $\myDetMat{D}_{\myKron}$ is a diagonal matrix whose diagonal entries are the eigenvalues of $\CovMat{\myKron}$ arranged in descending order. Let $\DaMat$ be an $\lenY\! \times \! \lenU^2$  diagonal matrix such that
	\vspace{-0.1cm}
	\begin{equation}
	\label{eqn:waterfil0}
	\left( \DaMat\right)_{k,k}^2 = \left(\tilde{\eta} - \frac{2\SigW}{\left(\myDetMat{D}_{\myKron}\right)_{k,k}  } \right)^+, \qquad \forall k \in \lenYSet, 
	\vspace{-0.1cm}
	\end{equation}
	 where $\tilde{\eta} $ is selected such that $\sum\limits_{k=1}^{\lenY} \big( \DaMat\big)_{k,k}^2\! =\! \frac{\myP^2}{\lenY}$. Finally, let
	 	\vspace{-0.1cm}
	 \begin{equation}
	 \label{eqn:waterfil1}
	 \tilde{\myA}\subopt = \DaMat  \myDetMat{V}_{\myKron}^H.
	 	\vspace{-0.1cm}
	 \end{equation} 
	 Then, $\tilde{\myA}\subopt$   satisfies the conditions in Lemma \ref{lem:optTildeA1}, computed with ${\myDetMat{E}}\big(\tilde{\myA}\subopt \big)$ replaced by ${\myDetMat{E}}_L\big(\tilde{\myA}\subopt \big)$.

%
%
\end{corollary} 
%
	{\em Proof:}
	See Appendix \ref{app:proofPro1}.
%

\smallskip
While Lemma \ref{lem:optTildeA1} corresponds to a generalized mercury waterfilling solution \cite[Thm. 4.2]{Carson:12}, Corollary \ref{pro:waterfil1} is reminiscent of the conventional waterfilling solution for the optimal $\tilde{\myA}$ when $\myKron$ is Gaussian \cite[Thm. 4.1]{Carson:12}. However, as noted in Subsection \ref{subsec:MIMORep},   $\myKron$ is non-Gaussian for any distribution of $\myU$, thus, the resulting $\tilde{\myA}\subopt$ has no claim of optimality.

\vspace{-0.2cm}
\subsection{Nearest Row-Wise Khatri-Rao Product Representation}
\label{subsec:AlgoPart3}
\vspace{-0.1cm}
The choice of $\tilde{\myA}\subopt$ in \eqref{eqn:waterfil1} does not necessarily correspond to a row-wise \ac{krp} structure \eqref{eqn:AtildeDef2}. In this case, it is not possible to find a matrix $\myA$ such that $|\myA \myU|^2 = \tilde{\myA}\subopt \left( \myU \otimes \myU^*\right)$, which implies that the matrix $\tilde{\myA}\subopt$ does not correspond to the model \eqref{eqn:model1}. 
Furthermore, we note that \ac{mi} is invariant to unitary transformations, and specifically, for any unitary $\myMat{V} \in \mySet{C}^{\lenY \times \lenY}$ and for any $\tilde{\myA} \in \mySet{C}^{\lenY \times \lenU^2}$ we have that 
	\vspace{-0.1cm}
\begin{align}
I\left(\myKron ; \tilde{\myA}\myKron \! + \! \myW_C \right) 
&\stackrel{(a)}{\! = \!} I\left(\myKron ; \tilde{\myA}\myKron \! + \! \myMat{V}^H\myW_C \right) \notag \\
&\stackrel{(b)}{\! = \!}I\left(\myKron ; \myMat{V}\tilde{\myA}\myKron \! + \! \myW_C \right),
	\vspace{-0.1cm}
\end{align}
where $(a)$ follows from \cite[Eq. (8.71)]{Cover:06}, and $(b)$ since $I\big(\myKron ;\myY_C\big) = I\big(\myKron ;\myMat{V}\myY_C\big)$, see \cite[Pg. 35]{Cover:06}.  
Therefore, in order to obtain a measurement matrix, we propose finding an $\lenY \times \lenU$ matrix $\hat{\myA}\nkrp$ such that, for a given unitary matrix $\myMat{V}$, 
\begin{equation}
	\label{eqn:KRPprobDef}
	\hat{\myA}\nkrp = \mathop {{\rm{argmin}}}\limits_{{\myA} \in \mySet{C}^{\lenY \times \lenU}} \|\myMat{V}\tilde{\myA}\subopt - \myS_{\lenY}\left(\myA \otimes \myA^* \right)\|^2.
\end{equation} 
Note that while the unitary matrix $\myMat{V}$ does not modify the \ac{mi},  it can result in reducing the minimal Frobenious norm in \eqref{eqn:KRPprobDef}. We will elaborate on the selection of $\myMat{V}$ in Subsection \ref{subsec:AlgoSummary}.

To solve \eqref{eqn:KRPprobDef}, let  $\tilde{\Avec}_k\subopt$ be the $\lenU^2 \times 1$ column vector corresponding to the $k$-th column of $\big( \myMat{V}\tilde{\myA}\subopt\big)^T$ and $\tilde{\myDetMat{M}}_{k}^{(H)}$ be the Hermitian part\footnote{The Hermitian part of a matrix $\myDetMat{Z}$ is given by $\frac{1}{2}\left(\myDetMat{Z}+\myDetMat{Z}^H \right)$.}  of ${\rm vec}_\lenU^{-1}\left( \tilde{\Avec}_{k}\subopt\right)$,  $k \in \lenYSet$.
The solution to \eqref{eqn:KRPprobDef} can be analytically obtained as stated in the following proposition: 
\begin{proposition}
	\label{pro:KRP}
	Let $\hat{\Avec}_k\nkrp$  be the $\lenU \times 1$ vector corresponding to the $k$-th column of $\big( \hat{\myA}\nkrp\big)^T$, $k \in \lenYSet$.
	Let $\tilde{\mu}_{k,\max}$  be the largest eigenvalue of $\tilde{\myDetMat{M}}_{k}^{(H)}$, and let $\tilde{\myVec{v}}_{k,\max}$ be the corresponding eigenvector, when the eigenvector matrix is unitary. Then, the columns of $\big( \hat{\myA}\nkrp\big)^T$ which solves  \eqref{eqn:KRPprobDef} are given by
	\vspace{-0.1cm}
	\begin{equation}
	\label{eqn:KRP}
	\hat{\Avec}_k\nkrp = \sqrt{\max\left( \tilde{\mu}_{k,\max},0\right) }\cdot  \tilde{\myVec{v}}_{k,\max}^*, \qquad k \in \lenYSet. 
	\vspace{-0.1cm}
	\end{equation}
%
%
\end{proposition}
%
	{\em Proof:}
	See Appendix \ref{app:proofKRP}.
%

\smallskip
The matrix $\hat{\myA}\nkrp$ derived in Proposition \ref{pro:KRP} does not necessarily satisfy the Frobenius norm constraint $\myP$. Thus, if the squared norm of  $\hat{\myA}\nkrp$ is larger than $\myP$, then it is scaled down to satisfy the norm constraint. 
Moreover, since $I\big(\myU ; \gamma |\hat{\myA}\nkrp \myU|^2 + \myW   \big)$ is monotonically non-decreasing w.r.t. $\gamma> 0$ \cite[Thm. 2]{Guo:05} for any distribution of $\myU$,  if the squared norm of  $\hat{\myA}\nkrp$ is smaller than $\myP$, then it is scaled up to the maximal norm to maximize the \ac{mi}. Consequently, the final measurement matrix is given by $\myA\nkrp = \frac{\sqrt{\myP}}{\left\|\hat{\myA}\nkrp\right\| }\hat{\myA}\nkrp$.

Next, we show that when $\myU$ is Kronecker symmetric, then, in the low \ac{snr} regime, $\myA\nkrp$ coincides with the optimal matrix characterized in Theorem \ref{thm:GaussianSource}, for any unitary transformation matrix $\myMat{V}$. 
Let $\myDetVec{i}_1$ be an $\lenY \times 1$ vector such that $\big(\myDetVec{i}_1\big)_k = \delta_{k,1}$, and let $\myDetMat{V}_{\myU}  \myDetMat{D}_{\myU}  \myDetMat{V}_{\myU}^H $ be the eigenvalue decomposition of $\CovMat{\myU}$. For a Kronecker symmetric $\myU$, we have that  $\CovMat{\myKron}=\CovMat{\myU}\otimes \CovMat{\myU}^*$, and thus $\myDetMat{V}_{\myKron}  = \myDetMat{V}_{\myU} \otimes \myDetMat{V}_{\myU}^*$ and $\myDetMat{D}_{\myKron}  = \myDetMat{D}_{\myU} \otimes \myDetMat{D}_{\myU}^*$ \cite[Ch. 12.3.1]{Golub:13}. 
\label{txt:Mod4}
\textcolor{NewColor}{
In the low \ac{snr} regime, due to the "waterfilling" in \eqref{eqn:waterfil1}, the measurement matrix extracts only  the least noisy spatial dimension of the \ac{soi}, resulting in $\tilde{\myA}\subopt = \frac{\myP}{\sqrt{\lenY}} \myDetVec{i}_1\big({\myVec{v}}_{\max} \otimes {\myVec{v}}_{\max}^*\big)^H$,  where ${\myVec{v}}_{\max}$ is the eigenvector corresponding to the maximal eigenvalue of the \ac{soi} covariance matrix, $\CovMat{\myU}$. }
Therefore, letting $\myDetVec{v}_1$ denote the leftmost column of   $\myMat{V}$, we have that $\myMat{V}\tilde{\myA}\subopt = \frac{\myP}{\sqrt{\lenY}} \myDetVec{v}_1\big({\myVec{v}}_{\max} \otimes {\myVec{v}}_{\max}^*\big)^H$, which results in ${\rm vec}_\lenU^{-1}\left( \tilde{\Avec}_{k}\subopt\right) = \frac{\myP}{\sqrt{\lenY}}\left( \myDetVec{v}_1\right)_k  {\myVec{v}}_{\max}{\myVec{v}}_{\max}^H$ \cite[Ch. 9.2]{Petersen:08} and $\tilde{\myDetMat{M}}_{k}^{(H)} = \frac{\myP}{\sqrt{\lenY}}{\rm Re} \left\{\left( \myDetVec{v}_1\right)_k \right\} {\myVec{v}}_{\max}{\myVec{v}}_{\max}^H$. Consequently, $\tilde{\myVec{v}}_{k,\max} =  {\myVec{v}}_{\max}$ for every $k \in \lenYSet$, and thus $\myA\nkrp$ is a rank-one matrix of the form $\myA\nkrp = \myDetVec{c} \cdot{\myVec{v}}_{\max}^H$, which coincides with $\myA\opt$ stated in  Theorem \ref{thm:GaussianSource}. For example, setting $\myMat{V} = \myI_\lenY$ results in $\myDetVec{c} = \sqrt{\myP} \cdot \myDetVec{i}_1$. 

\vspace{-0.2cm}
\subsection{Masked Fourier Measurement Matrix}
\label{subsec:AlgoFourier} 
\vspace{-0.1cm}
As mentioned in Subsection \ref{subsec:Pre_Model}, in many phase retrieval setups, the measurement matrix represents masked Fourier measurements and is constrained to the structure of \eqref{eqn:KRP2AMatForm}. 
In the context of phase retrieval, the design goal is to find the set of masks $\{\Gmat_l\}_{l=1}^{b}$ in \eqref{eqn:KRP2AMatForm} which result in  optimal recovery performance. To that aim, define the $\lenU \times 1$ vectors $\gvec_l$, $l \in  \mathcal{B}$, to contain the diagonal elements of  $\Gmat_{l} $, $\left( \gvec_l\right)_k \! =\! \left(\Gmat_{l} \right)_{k,k}$, $k \in \lenUSet$. With this definition, we can write  
\vspace{-0.1cm}
\begin{equation}
\label{eqn:MaskedFMatDef2}
\left(\myA \right)_{(l\!-\!1)\lenU + k, p} \!=\! \left( \gvec_l\right)_p  \left( \Fmat{\lenU}\right)_{k,p}, \quad  \forall k, p  \!\in \!\lenUSet, l\! \in \! \mathcal{B}.
\vspace{-0.1cm}
\end{equation}
Since ${\myA}\nkrp$  does not necessarily represent a masked Fourier structure, based on the rationale detailed in Subsection \ref{subsec:AlgoPart3}, we suggest to use the masks $\{\gvec_l\maskedF\}_{l=1}^{b}$ that minimize the distance between the resulting measurement matrix and a unitary transformation of $\tilde{\myA}\subopt$:
\vspace{-0.15cm}
\begin{equation}
\label{eqn:KRP2probDef}
\!\!\!\{\gvec_l\maskedF\}_{l\! = \!1}^{b} \! = \!\mathop {{\rm{argmin}}}\limits_{\{\gvec_l\}_{l=1}^{b}\in \mySet{C}^{\lenU}} \|\myMat{V}\tilde{\myA}\subopt \! - \! \myS_{\lenY}\left(\myA \otimes \myA^* \right)\|^2,
\vspace{-0.1cm}
\end{equation}
where $\myMat{V}$ is a given unitary matrix and $\myA$ depends on $\{\gvec_l\maskedF\}_{l=1}^{b}$ via \eqref{eqn:MaskedFMatDef2}. 
The set of masks which solve \eqref{eqn:KRP2probDef} is characterized in the following proposition: 
\begin{proposition}
	\label{pro:KRP2}
	Let $\tFmat{k}$ be an $\lenU \times \lenU$ diagonal matrix such that  $\big(\tFmat{k}  \big)_{p,p} \!=\! \big( \Fmat{\lenU}\big)_{k,p}$, $k,p \!\in\! \lenUSet$.  
	For all $l \!\in\! \mathcal{B}$, let $\bar{\mu}_{l,\max}$  be the largest eigenvalue of the $\lenU \times \lenU$ Hermitian matrix  $\sum\limits_{k=1}^{\lenU}\tFmat{k} \tilde{\myDetMat{M}}_{(l-1)\lenU+k}^{(H)}\tFmat{k}^*$, where  $\tilde{\myDetMat{M}}_{(l-1)\lenU+k}^{(H)}$ is the Hermitian part of ${\rm vec}_\lenU^{-1}\left( \tilde{\Avec}_{(l-1)\lenU+k}\subopt\right)$, 
	and let $\bar{\myVec{v}}_{l,\max}$   be its corresponding eigenvector, when the eigenvector matrix is unitary. 
	 Then, the set of mask coefficients $\{\gvec_l\maskedF\}_{l=1}^{b}$ which solves \eqref{eqn:KRP2probDef} is obtained as
	 \vspace{-0.1cm}
	\begin{equation}
	\label{eqn:KRP2}
	\gvec_l\maskedF = \sqrt{\lenU \cdot \max\left(\bar{\mu}_{l,\max},0\right)} \cdot \bar{\myVec{v}}_{l,\max}^*, \qquad l \in \mathcal{B}. 
	\vspace{-0.1cm}
	\end{equation}
%
%
\end{proposition} 
%
{\em Proof:}
	See Appendix \ref{app:proofKRP2}.
%

\smallskip
The masked Fourier measurement matrix is obtained from the coefficient vectors $\{\gvec_l\maskedF\}_{l\! = \!1}^{b}$  via 
\vspace{-0.1cm}
\begin{equation}
\label{eqn:MaskedRes1}
\!\big( \hat{\myA}\maskedF\big)_{(l\!-\!1)\cdot\lenU\! +\! k, p} \!=\! \left( \gvec_l\maskedF \right)_p  \left( \Fmat{\lenU}\right)_{k,p}, \!\! \quad k, p  \!\in\! \lenUSet, \; l\! \in\!  \mathcal{B}.
\vspace{-0.1cm}
\end{equation}
Applying the same reasoning used in determining the scaling of $\hat{\myA}\nkrp$ in Subsection \ref{subsec:AlgoPart3}, we conclude that the \ac{mi} is maximized, subject to the trace constraint, by normalizing $\hat{\myA}\maskedF$ to obtain
 $\myA\maskedF = \frac{\sqrt{\myP}}{\left\|\hat{\myA}\maskedF\right\| }\hat{\myA}\maskedF$.
   
Let us again consider a Kronecker symmetric $\myU$ in the low \ac{snr} regime. 
For simplicity, we set $\myMat{V} = \myI_\lenY$. 
\label{txt:Mod5}
\textcolor{NewColor}{
As discussed in the previous subsection, for this setting we have that $\tilde{\myA}\subopt = \frac{\myP}{\sqrt{\lenY}} \myDetVec{i}_1\big(\myVec{v}_{\max} \otimes \myVec{v}_{\max}^*\big)^H$, where $\myDetVec{i}_1$ is the $\lenY \times 1$ vector such that $\big(\myDetVec{i}_1\big)_k = \delta_{k,1}$,} 
and thus $\tilde{\myDetMat{M}}_{k}^{(H)}$ is non-zero only for $k=1$. Therefore, $\bar{\mu}_{l,\max}$ is zero for all $l \neq 1$, while $\bar{\mu}_{1,\max}$ is the largest eigenvalue of $\tFmat{1}^* \tilde{\myDetMat{M}}_{1}^{(H)}\tFmat{1} = {\myDetMat{M}}_{1}^{(H)} =   \myVec{v}_{\max}\myVec{v}_{\max}^H$, and thus $\bar{\myVec{v}}_{1,\max} = \myVec{v}_{\max}$. Consequently, we have that 
\begin{equation}
\label{eqn:KRP2AMatForm2}
\myA\maskedF 
= \sqrt{\myP}\left[ {\begin{array}{*{20}{c}}
	\Fmat{\lenU}{\rm diag}\left( \myVec{v}_{\max}^*\right)  \\
	0\quad \ldots \quad 0 \\
	\vdots \\
	0\quad \ldots \quad 0 
	\end{array}} \right].
\vspace{-0.1cm}
\end{equation}
Unlike the unconstrained case considered in the previous subsection, the resulting measurement matrix in \eqref{eqn:KRP2AMatForm2} does not coincide with the optimal matrix given in Theorem \ref{thm:GaussianSource}, due to the masked Fourier structure constraint. 

\vspace{-0.2cm}
\subsection{Obtaining the Optimal Unitary Transformation Matrix}
\label{subsec:AlgoSummary}
\vspace{-0.1cm}
In the previous subsections we assumed that the unitary transformation $\myMat{V}$ applied to $\tilde{\myA}\subopt$ is given. In the following we propose an algorithm to jointly identify the optimal transformation $\myMat{V}$ and the optimal measurement matrix $\myA$.  

Let $\mySet{V}$ denote the set of $\lenY \times \lenY$ complex unitary matrices and $\mySet{A}$ denote the set of $\lenY \times \lenU$ feasible measurement matrices. For example, for unconstrained measurements, $\mySet{A}=\mySet{C}^{\lenY \times \lenU}$, and for masked Fourier measurements,  $\mySet{A}$ is the set of all matrices which can be expressed as in \eqref{eqn:KRP2AMatForm}. The optimal $\myA$ and $\myMat{V}$ are obtained as the solution to the following joint optimization problem:
\vspace{-0.15cm}
\begin{equation}
\label{eqn:KRP2probDef2}
\!\!\!\Big( \hat{\myA}\maskedFU ,\myMat{V}\maskedFU \Big)   \! = \!\mathop {{\rm{argmin}}}\limits_{\myA \in \mySet{A}, \myMat{V} \in \mySet{V}} \|\myMat{V} \tilde{\myA}\subopt \! - \! \myS_{\lenY}\left(\myA \otimes \myA^* \right)\|^2.
\vspace{-0.1cm}
\end{equation}
%
The solution to \eqref{eqn:KRP2probDef2} {\em for a fixed $\myMat{V}$} is given in Propositions \ref{pro:KRP} and \ref{pro:KRP2}. 
For a fixed $\myA$, the problem in \eqref{eqn:KRP2probDef2} is the unitary Procrustes problem \cite[Ch. 7.4]{Horn:85}: Letting $\myDetMat{V}_{\rm svd}\left(\myA \right)\myDetMat{D}_{\rm svd}\left(\myA \right)\myDetMat{W}_{\rm svd}^H\left(\myA \right)$ be the \ac{svd} of $\myS_{\lenY}\left(\myA \otimes \myA^* \right) \cdot \big(\tilde{\myA}\subopt\big)^H$, the solution to \eqref{eqn:KRP2probDef2}   {\em for a fixed $\myA$} is given by 
\begin{equation}
\label{eqn:Procrustes1}
\myMat{V}\maskedFU \left(\myA \right) =  \myDetMat{V}_{\rm svd}\left(\myA \right)\myDetMat{W}_{\rm svd}^H\left(\myA \right).
\end{equation}

Based on the above, we propose to solve the joint optimization problem \eqref{eqn:KRP2probDef2} in an alternating fashion, i.e., optimize over $\mySet{A}$ for a fixed $\myMat{V}$,  then optimize over $\mySet{V}$ for a fixed $\myA$,  and continue with the alternating optimization process until convergence. The overall matrix design algorithm is summarized in  Algorithm \ref{alg:Algo1}. 
{As the Frobenious norm objective in \eqref{eqn:KRP2probDef2} is differentiable, convergence of the alternating optimization algorithm is guaranteed \cite[Thm. 2]{Bezdek:03}. However, since the problem is not necessarily convex\footnote{This non-convexity is observed by noting that, for example, for $\phi \in (0,2\pi)$, the right hand side of \eqref{eqn:KRP2probDef2} obtains the same value for $\myA$ and for $\myA e^{j\phi}$, and a different value for $\frac{1}{2}(1 +  e^{j\phi})\myA$, which is an element of every convex set containing $\myA$ and $\myA e^{j\phi}$. Consequently, when $\myA$ which is not all zero solves \eqref{eqn:KRP2probDef2}, the set of all minima is not convex, and the optimization problem is thus not convex \cite[Ch. 4.2]{Boyd:04}.} w.r.t. both $\myA$ and $\myMat{V}$, the algorithm may converge to a local minima.}

%
\begin{algorithm}
	\caption{Measurement Matrix Design}
	\label{alg:Algo1}
	\begin{algorithmic}[1]
		\STATE Initialization: Set $k=0$ and $\myMat{V}_0 = \myMat{I}_\lenY$.
		\STATE \label{stp:MF0} Compute $\tilde{\myA}\subopt$ using \eqref{eqn:waterfil1}.
		\STATE Obtain  $\hat{\myA}_{k\!+\!1}\!=\!\mathop {{\rm{argmin}}}\limits_{\myA \in \mySet{A}} \label{stp:MF1} \|\myMat{V}_k \tilde{\myA}\subopt \! - \! \myS_{\lenY}\left(\myA \otimes \myA^* \right)\|^2$ using  Proposition \ref{pro:KRP} (for general measurements) or using Proposition \ref{pro:KRP2} (for masked Fourier measurements).
		\STATE \label{stp:MF2} Set  $\myMat{V}_{k\!+\!1}\!=\! \myDetMat{V}_{\rm svd}\big(\hat{\myA}_{k\!+\!1}\big)\myDetMat{W}_{\rm svd}^H\big(\hat{\myA}_{k\!+\!1}\big)$.
		\STATE If termination criterion is inactive: Set $k := k+1$ and go to Step \ref{stp:MF1}.	
		\STATE  $\myA\maskedFU$ is obtained as   $\myA\maskedFU = \frac{\sqrt{\myP}}{\left\|\hat{\myA}_{k}\right\| }\hat{\myA}_{k}$.
\end{algorithmic}
\end{algorithm}

\label{txt:Complexity1}
\textcolor{NewColor}{
Assuming that the computation of $\tilde{\myA}\subopt$ in Step \ref{stp:MF0} of Algorithm \ref{alg:Algo1} is carried out using a computationally efficient waterfilling algorithm, as in, e.g., \cite{Palomar:05}, the  complexity of Algorithm \ref{alg:Algo1} is dominated by the computation of the eigenvalue decomposition required in Step \ref{stp:MF0} and by the matrix product required to compute the \ac{svd} in Step \ref{stp:MF2}. Letting $t_{\max}$ denote the maximal number of iterations over Steps \ref{stp:MF1}-\ref{stp:MF2}, it follows that the   the overall computational complexity of the algorithm is on the order of $\mathcal{O}(t_{\max}\cdot \lenY^2 \cdot \lenU^2 + \lenU^6 )$\cite[Ch. 1.1, Ch. 8.6]{Golub:13}.}

\label{txt:Colored1}
\textcolor{NewColor}{
While in the problem formulation we consider white Gaussian noise, the measurement matrix design in Algorithm \ref{alg:Algo1} can be extended to account for colored Gaussian noise, i.e., for noise  $\myW$ with covariance matrix  $\CovMat{\myW} \neq \SigW \myI_{\lenY}$, by considering the whitened observations vector $\CovMat{\myW}^{-1/2}\myY$ instead of $\myY$. This is because invertible transformations do not change the \ac{mi}: $I\left(\myU; \myY \right) =  I\left(\myU; \CovMat{\myW}^{-1/2}\myY \right)$ \cite[Corollary after Eq. (2.121)]{Cover:06}, therefore maximizing the \ac{mi} for the whitened observations maximizes the \ac{mi} for the original observations. After applying the whitening transformation, Algorithm \ref{alg:Algo1} can be used on the whitened observations vector $\CovMat{\myW}^{-1/2}\myY$ with noise covariance  matrix $\myI_{\lenY}$, with the exception that the objective function in  Step \ref{stp:MF1} is replaced with $\mathop {{\rm{argmin}}}\limits_{\myA \in \mySet{A}} \|\myMat{V}_k \CovMat{\myW}^{1/2}\tilde{\myA}\subopt \! - \! \myS_{\lenY}\left(\myA \otimes \myA^* \right)\|^2$. 
}

\vspace{-0.2cm}
\section{Simulations Study}
\label{sec:Simulations}
\vspace{-0.1cm}
In this section we evaluate the performance of phase retrieval with the proposed measurement matrix design in a simulations study. 
While our design aims at {maximizing the statistical dependence between the \ac{soi} and the observations via \ac{mi} maximization}, we note that phase retrieval is essentially {\em an estimation problem}, hence, we evaluate the performance in terms of estimation error. Since  the phase retrieval setup inherently has a global phase ambiguity, for an \ac{soi} realization $\myU \!=\! \myDetVec{u}$ and its  estimate $\hat{\myU} \!=\! \hat{\myDetVec{u}}$, we define the estimation error  as 
\vspace{-0.1cm}
\begin{equation}
\epsilon\left( \myDetVec{u}, \hat{\myDetVec{u}} \right) = \mathop{\min}\limits_{c \in \mySet{C} : |c| = 1} \frac{\|\myDetVec{u} - c\cdot \hat{\myDetVec{u}} \|}{\|\myDetVec{u}\|}, 
\label{eqn:ErrorDef1}
\vspace{-0.1cm}
\end{equation}
namely, the minimum relative distance over all phase rotations, see, e.g., \cite[Eq. (19)]{Waldspruger:15}.
We use both phasecut \cite{Waldspruger:15} and \ac{taf} (with step-size $1$ and truncation threshold $0.9$) \cite{Wang:16} to estimate the \ac{soi} $\myU$ from the observations $\myY$. 
%
Performance was evaluated for five different measurement matrices:
\begin{itemize}
	\item  $\myA^{\rm OK}$ - The optimal measurement matrix for Kronecker symmetric \ac{soi} in the low \ac{snr} regime, obtained via \eqref{eqn:GaussianSource} with  $\myDetVec{c}$ selected such that $\left( \myDetVec{c}\right)_k = \sqrt{\frac{\myP}{\lenY}}e^{j2\pi\frac{k-1}{\lenY}}$ for all $k \in \lenYSet$.
	\item $\myA^{\rm UC}$  - The unconstrained measurement matrix obtained using Algorithm \ref{alg:Algo1} with $\mySet{A} = \mySet{C}^{\lenY\times\lenU}$.
	\item $\myA^{\rm MF}$  - The masked Fourier measurement matrix obtained using Algorithm \ref{alg:Algo1} with $\mySet{A}$ being the set of matrices which can be expressed as in \eqref{eqn:KRP2AMatForm}.
	\item  $\myA^{\rm RG}$  - A random \ac{pc} Gaussian matrix with i.i.d. entries.
	\item  $\myA^{\rm CD}$  - A coded diffraction pattern matrix with random octanary patterns \cite[Sec. 4.1]{Candes:15a}, namely, a masked Fourier matrix \eqref{eqn:KRP2AMatForm} with i.i.d. random masks, each having i.i.d. entries distributed according to~\cite[Eq. (4.3)]{Candes:15a}. 
\end{itemize}
For the random matrices, $\myA^{\rm RG}$ and $\myA^{\rm CD}$,  a new realization is generated for each Monte Carlo simulation. 
The squared Frobenius norm constraint is set to $\myP\! = \!\lenY$, namely, the average  row squared norm for all designed matrices is $1$. 
Two different \ac{soi} distributions of size $\lenU = 10$ were tested:
\begin{itemize}
	\item $\myU_S$ - A sum of complex exponentials (see, e.g., \cite[Sec. V]{Waldspruger:15}) given by $\left( \myU_S\right)_k  = \sum\limits_{l=1}^6 M_l e^{j \pi \Phi_l k}$, where $\{ M_l \}_{l=1}^6$ are i.i.d. zero-mean unit variance real-valued  Gaussian \ac{rv}s, and $\{ \Phi_l \}_{l=1}^6$ are i.i.d. \ac{rv}s uniformly distributed over $[0,\pi]$, independent of $\{ M_l \}_{l=1}^6$.	
	\item $\myU_G$ -  A zero-mean \ac{pc} Gaussian vector with covariance matrix $\CovMat{\myU}$ corresponding to an exponentially decaying correlation profile given by $\left( \CovMat{\myU}\right)_{k,l} = 6\cdot e^{-|k-l| +j\frac{2\pi(k-l)}{\lenU}}$, $k,l \in \lenUSet$. 
\end{itemize}
Note that all tested \ac{soi}s have the same energy, measured as the trace of the covariance matrix. 
%
The estimation error is averaged over $1000$ Monte Carlo simulations, where a new \ac{soi} and noise realization is generated in each simulation. 

\begin{figure}
	\centering
	\vspace{-0.2cm}
	\scalebox{0.5}
	{\includegraphics{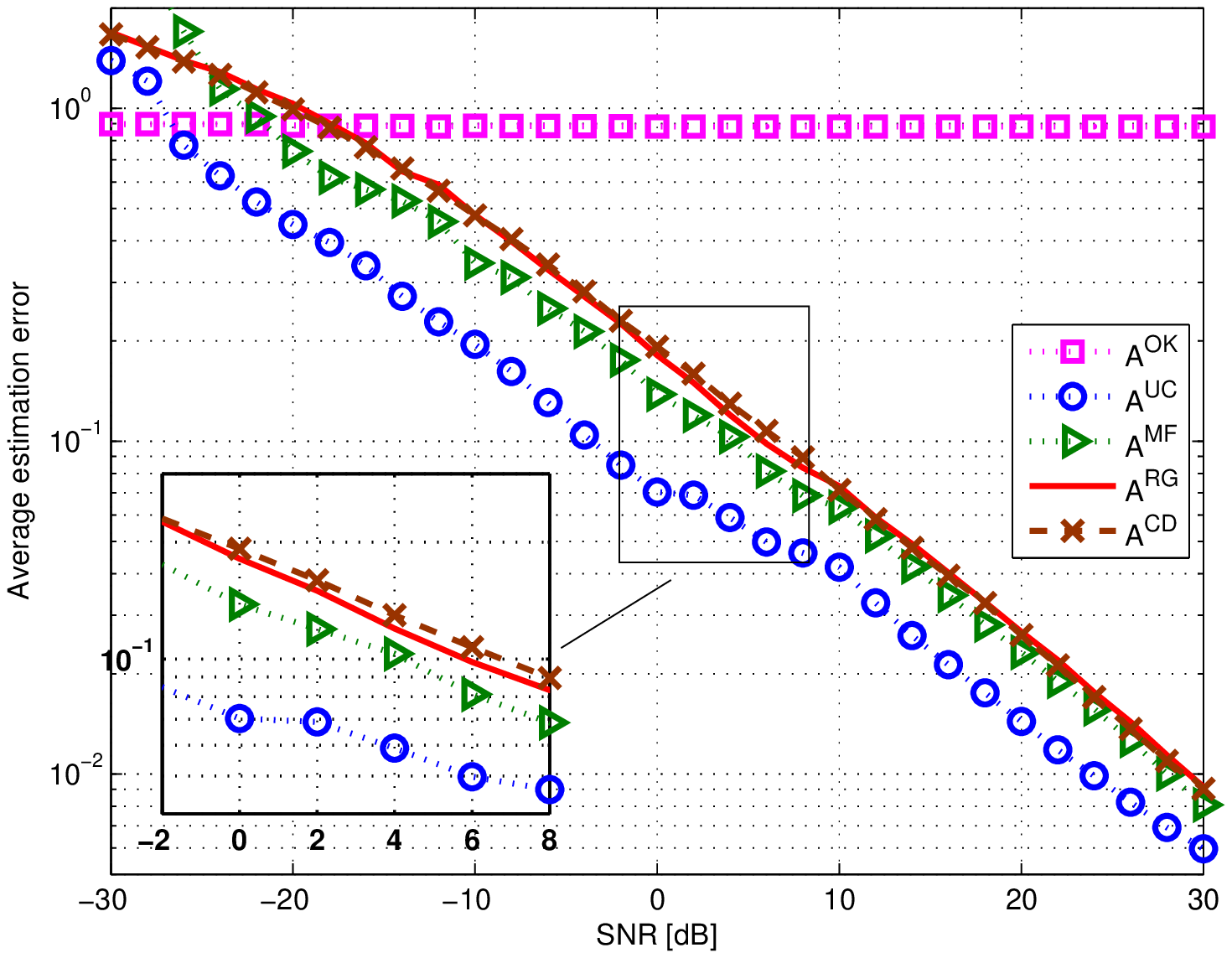}}
	\vspace{-0.6cm}
	\caption{Average estimation error vs. \ac{snr} for $\myU_S$ using phasecut, $\lenY = 6 \lenU$.}
	\vspace{-0.2cm}
	\label{fig:NumEx1a}
\end{figure}

\begin{figure}
	\centering
	\vspace{-0.2cm}
	\scalebox{0.5}
	{\includegraphics{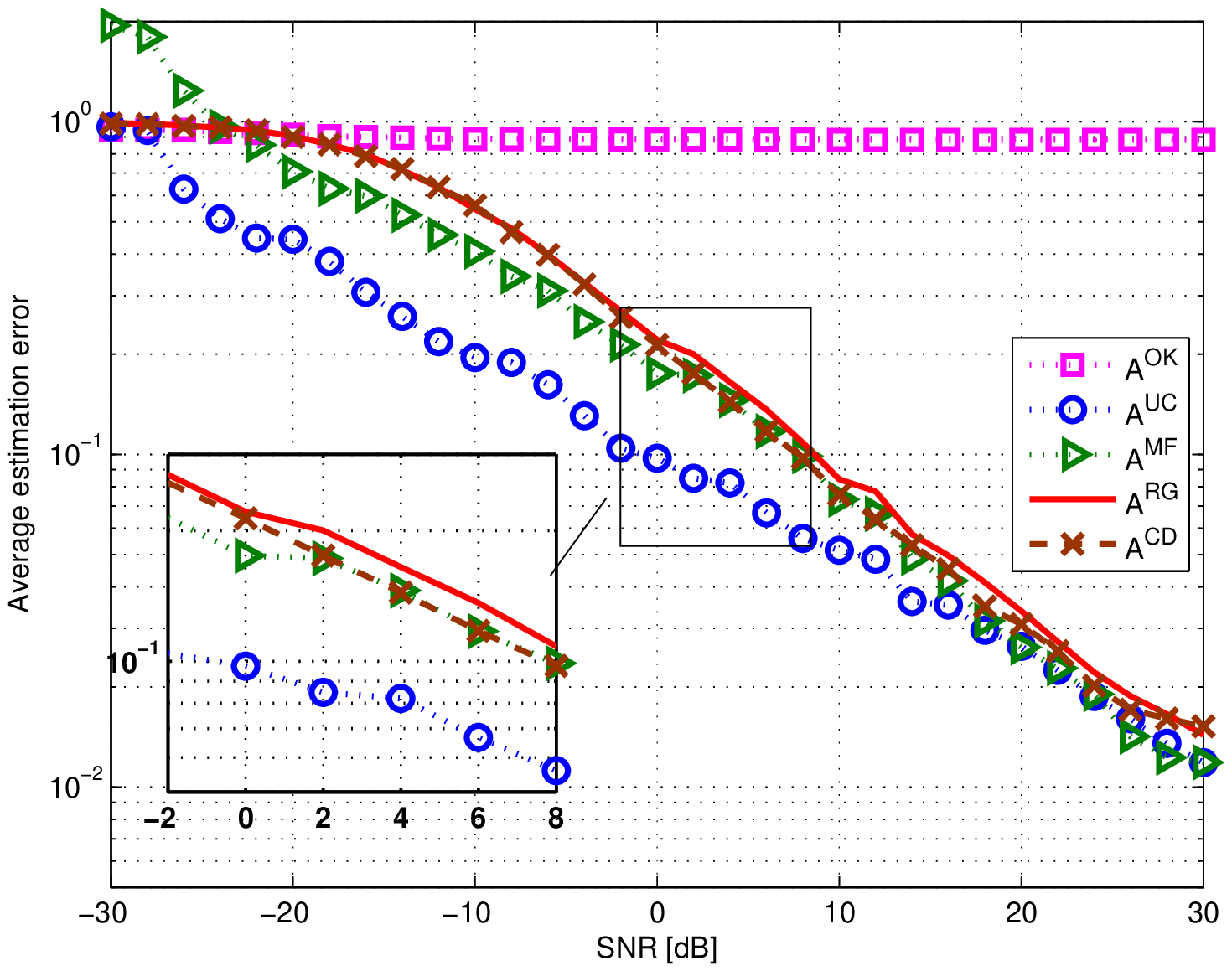}}
	\vspace{-0.6cm}
	\caption{Average estimation error vs. \ac{snr} for $\myU_S$ using \ac{taf}, $\lenY = 6 \lenU$.}
	\vspace{-0.2cm}
	\label{fig:NumEx1a2}
\end{figure}

\begin{figure}
	\centering
	\vspace{-0.2cm}
	\scalebox{0.5}
	{\includegraphics{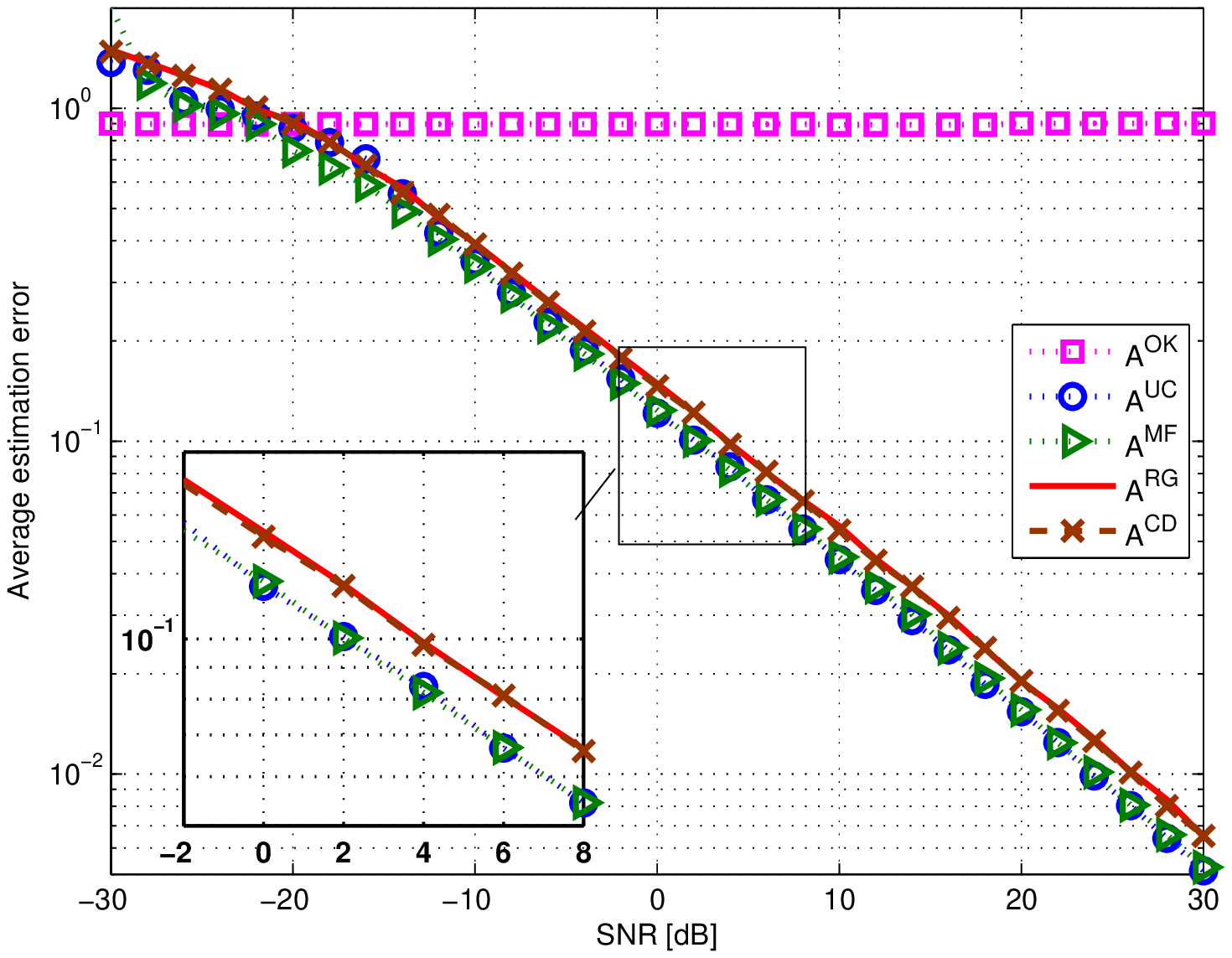}}
	\vspace{-0.6cm}
	\caption{Average estimation error vs. \ac{snr} for $\myU_G$ using  phasecut, $\lenY = 6 \lenU$.}
	\vspace{-0.3cm}
	\label{fig:NumEx1b}
\end{figure}

\begin{figure}
	\centering
	\vspace{-0.2cm}
	\scalebox{0.5}
	{\includegraphics{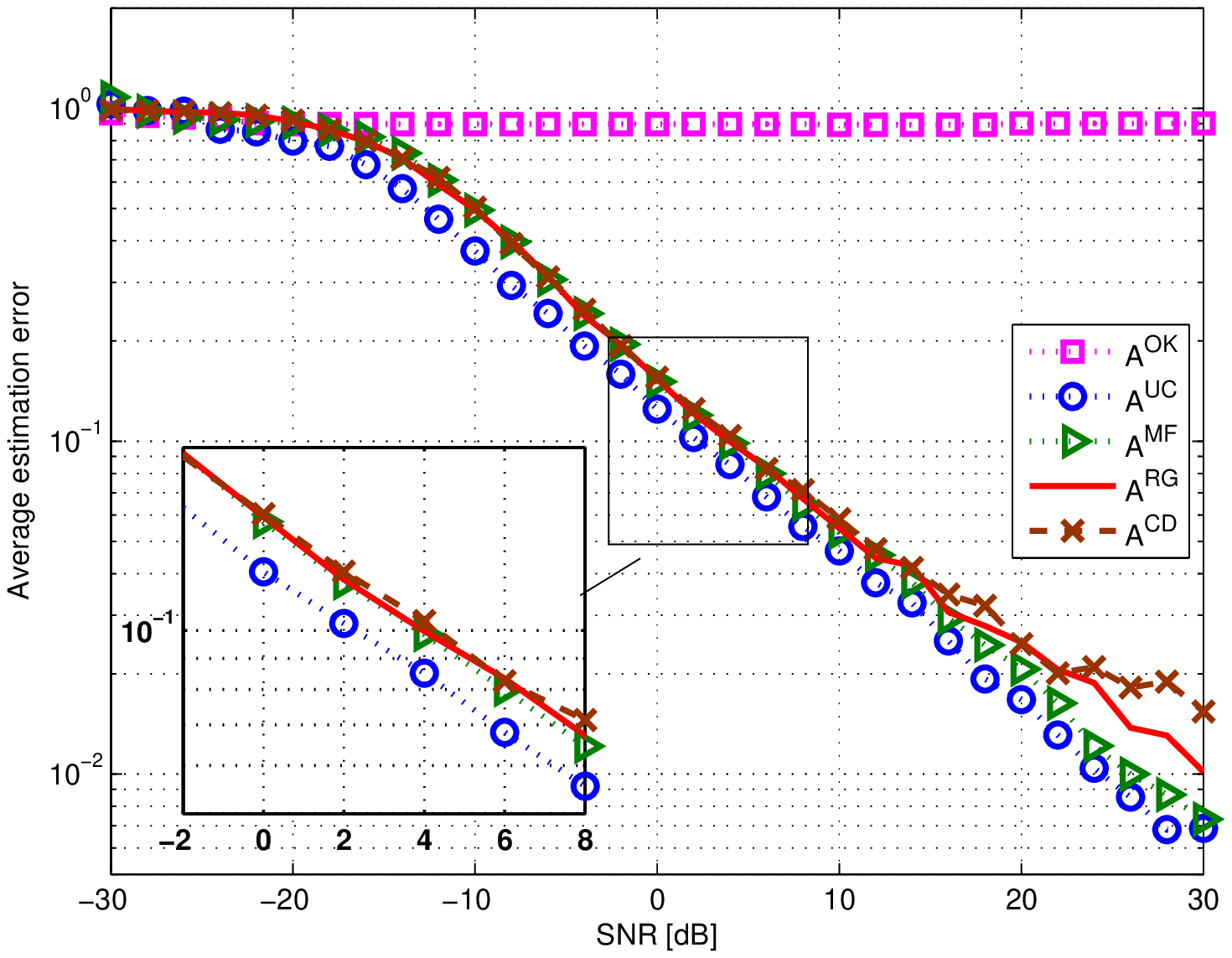}}
	\vspace{-0.6cm}
	\caption{Average estimation error vs. \ac{snr} for $\myU_G$ using \ac{taf}, $\lenY = 6 \lenU$.}
	\vspace{-0.2cm}
	\label{fig:NumEx1b2}
\end{figure}

In Figs. \ref{fig:NumEx1a}--\ref{fig:NumEx1b2} we fix the observations dimension  to be $\lenY = 6 \cdot \lenU = 60$, and let the \ac{snr}, defined as $1/{\SigW}$, vary from $-30$ dB to $30$ dB, for $\myU_S$ using phasecut, $\myU_S$ using \ac{taf}, $\myU_G$ using phasecut, and $\myU_G$ using \ac{taf}, respectively. 
%
%
It can be observed from Figs. \ref{fig:NumEx1a}--\ref{fig:NumEx1b2} that the {\em deterministic unconstrained}  $\myA^{\rm UC}$ achieves the best performance over almost the entire \ac{snr} range, for all tested \ac{soi} distributions. Notable gains are observed for  $\myU_S$ in Figs. \ref{fig:NumEx1a}--\ref{fig:NumEx1a2}, where, for example, $\myA^{\rm UC}$ attains an average estimation error of $\epsilon \!=\! 0.1$ for \ac{snr}s of $-4$ dB and $-2$ dB, for phasecut and for \ac{taf}, respectively, while random Gaussian measurements $\myA^{\rm RG}$  achieve $\epsilon \!=\! 0.1$ for \ac{snr}s of $4$ dB and $8$ dB, for phasecut and for \ac{taf}, respectively,  and random coded diffraction patterns $\myA^{\rm CD}$  achieve $\epsilon \!=\! 0.1$ for \ac{snr}s of $6$ dB and $8$ dB, for phasecut and for \ac{taf}, respectively. Consequently, for \ac{soi} distribution $\myU_S$, $\myA^{\rm UC}$ achieves an \ac{snr} gain of $8-10$ dB at $\epsilon \!= \!0.1$ over Gaussian measurements, and an \ac{snr} gain of $10$ dB over random coded diffraction patterns.  From Figs. \ref{fig:NumEx1b}--\ref{fig:NumEx1b2} we observe that  the corresponding \ac{snr} gain at $\epsilon\! =\! 0.1$ for the \ac{soi} distribution $\myU_G$ is $2$ dB, compared to both random Gaussian measurements as well as to  random coded diffraction patterns. 
Furthermore, it is observed from  Figs. \ref{fig:NumEx1a}--\ref{fig:NumEx1b2} that the proposed masked Fourier measurement matrix $\myA^{\rm MF}$, corresponding to {\em practical deterministic masked Fourier measurements}, achieves an \ac{snr} gain of $0-2$ dB  for both \ac{soi} distributions $\myU_G$  and $\myU_S$, compared to random Gaussian measurements and random coded diffraction patterns. 
%
It is also noted in Figs. \ref{fig:NumEx1a}--\ref{fig:NumEx1b2} that, as expected, in the low \ac{snr} regime, i.e.,   $1 / \SigW < -20$ dB, $\myA^{\rm OK}$ obtains the best performance, as it is designed specifically for low \ac{snr}s. However, the performance of  $\myA^{\rm OK}$  for both recovery algorithms hardly improves with \ac{snr} as its rank-one structure does not allow the complete recovery of the \ac{soi} at any \ac{snr}. 

\begin{figure}
	\centering
	\vspace{-0.2cm}
	\scalebox{0.5}
	{\includegraphics{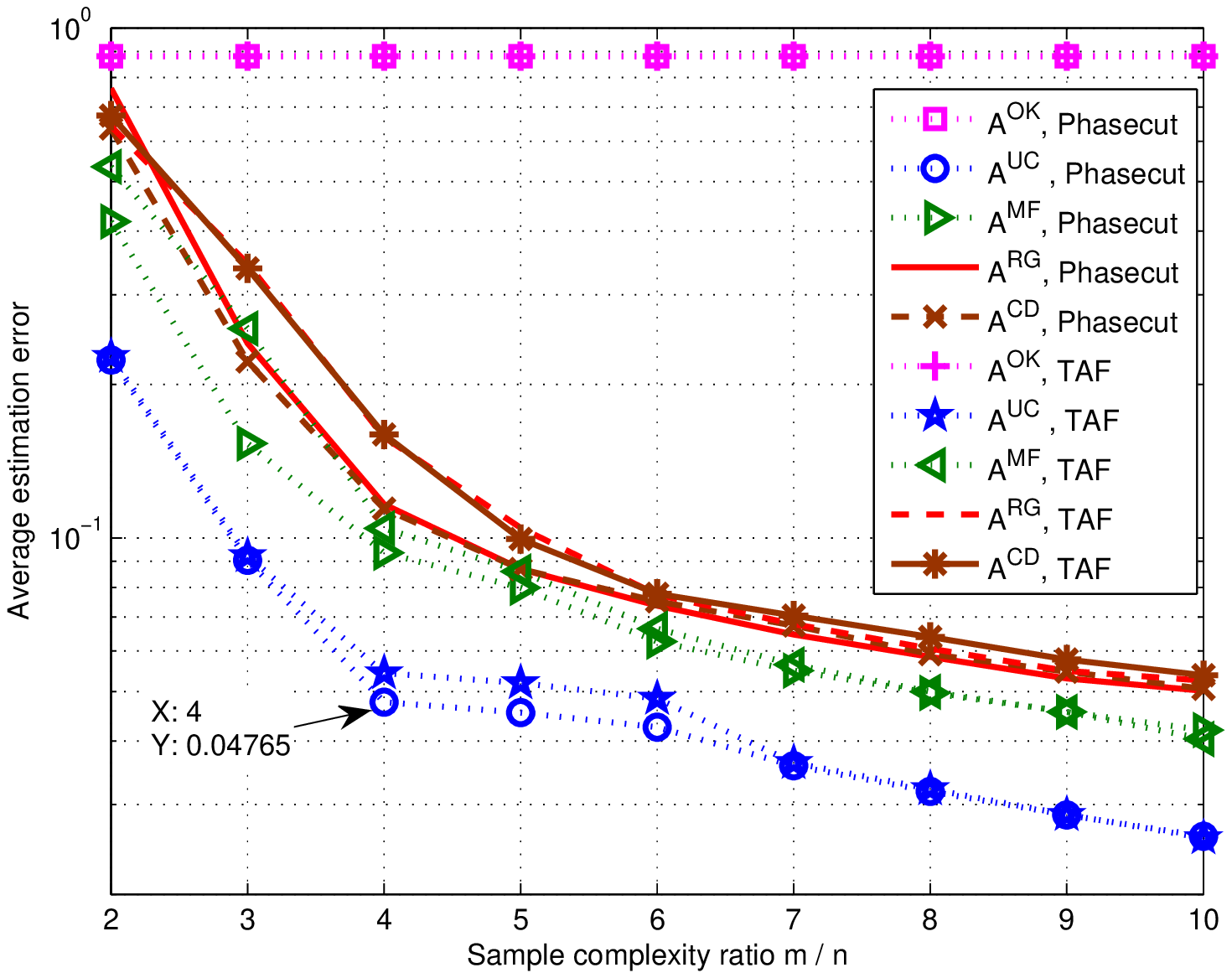}}
	\vspace{-0.6cm}
	\caption{Average estimation error vs. sample complexity, $\myU_S$,  \ac{snr} $\!=\!10$ dB.}
	\vspace{-0.2cm}
	\label{fig:NumEx3a}
\end{figure}

\begin{figure}
	\centering
	\vspace{-0.2cm}
	\scalebox{0.5}
	{\includegraphics{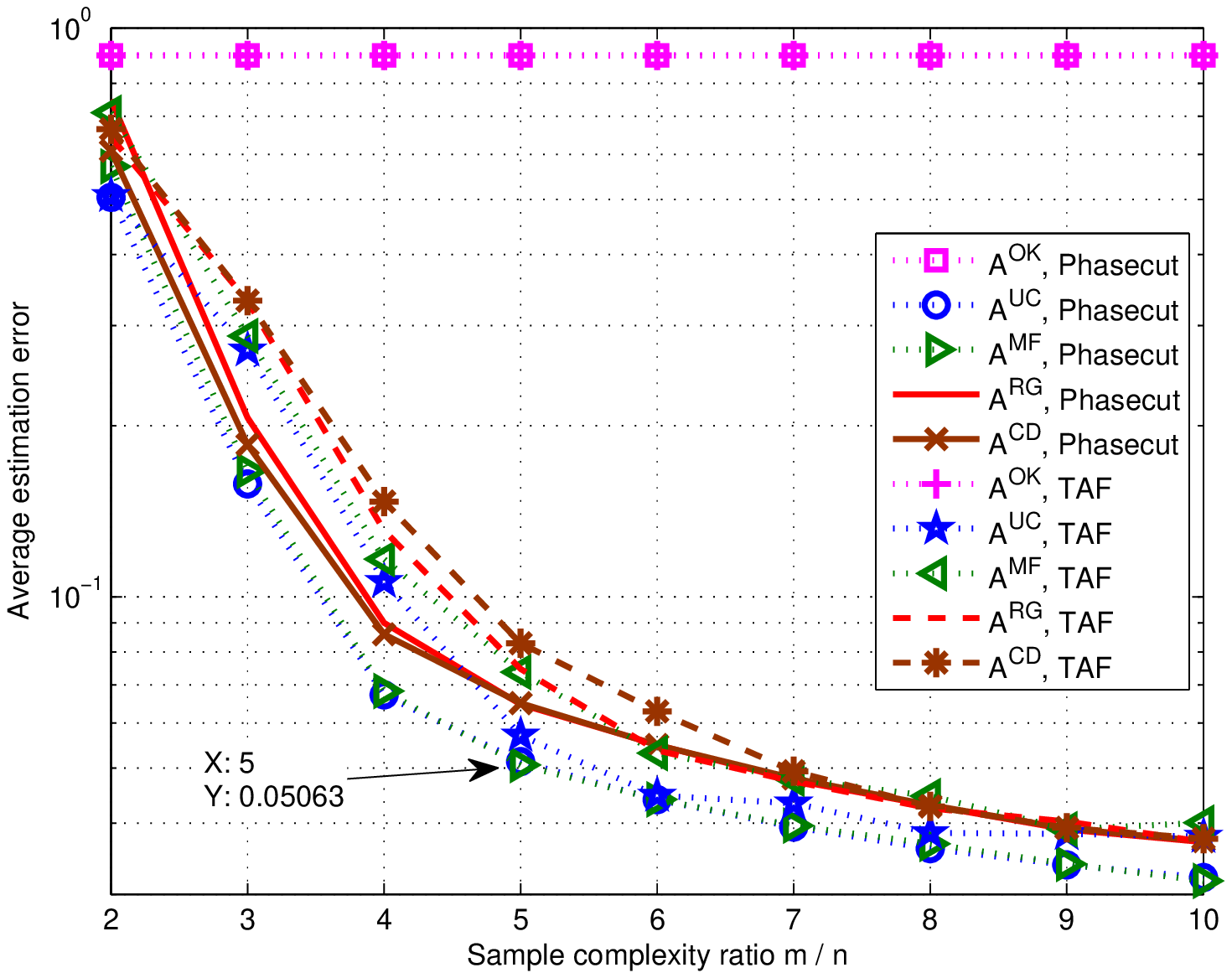}}
	\vspace{-0.6cm}
	\caption{Average estimation error vs. sample complexity, $\myU_G$,  \ac{snr} $\!=\!10$ dB.}
	\vspace{-0.2cm}
	\label{fig:NumEx3b}
\end{figure}

In Figs. \ref{fig:NumEx3a}--\ref{fig:NumEx3b} we fix the \ac{snr}  to be $10$ dB, and let the sample complexity ratio $\frac{\lenY}{\lenU}$ \cite{Candes:15a, Wang:16} vary from $2$ to $10$, for both $\myU_S$ and $\myU_G$.
From Figs. \ref{fig:NumEx3a}-\ref{fig:NumEx3b} we observe that the superiority of the deterministic $\myA^{\rm UC}$ is maintained for different sample complexity values. For example, in Fig. \ref{fig:NumEx3a} we observe that for $\myU_S$ at \ac{snr} $1/\SigW = 10$ dB,  $\myA^{\rm UC}$ obtains an estimation error of less than $\epsilon = 0.05$ for $\lenY = 4\lenU$ and for $\lenY = 6\lenU$, using phasecut and using \ac{taf}, respectively, while  our masked Fourier design $\myA^{\rm MF}$ requires $\lenY = 8\lenU$ observations, and both random Gaussian measurements and random coded diffraction patterns require $\lenY = 10\lenU$ observations to achieve a similar estimation error, for both phasecut and \ac{taf}. 
A similar behavior with less notable gains is observed for  $\myU_G$  in Fig.  \ref{fig:NumEx3b}. For example, for $\myU_G$ using phasecut, both $\myA^{\rm UC}$ and $\myA^{\rm MF}$  require $\lenY = 5\lenU$ observations to achieve $\epsilon = 0.05$, while  both $\myA^{\rm RG}$ and $\myA^{\rm CD}$ require $\lenY = 7\lenU$ observations to achieve similar performance. 
This implies that our proposed designs {\em require fewer measurements}, compared to the common random measurement matrices, to achieve the same performance. 

%

Moreover, we observe that the estimation error of both the unconstrained measurements $\myA^{\rm UC}$ and  the masked Fourier measurements ${\myA}^{\rm MF}$ scale w.r.t. \ac{snr} (Figs. \ref{fig:NumEx1a}--\ref{fig:NumEx1b2}) and sample complexity (Figs. \ref{fig:NumEx3a}--\ref{fig:NumEx3b}) similarly to random measurements  $\myA^{\rm RG}$ and $\myA^{\rm CD}$, and that the performance gain compared to random Gaussian measurements and random coded diffraction patterns is maintained for various values of $m$.

\begin{figure}
	\centering
	\vspace{-0.2cm}
	\scalebox{0.5}
	{\includegraphics{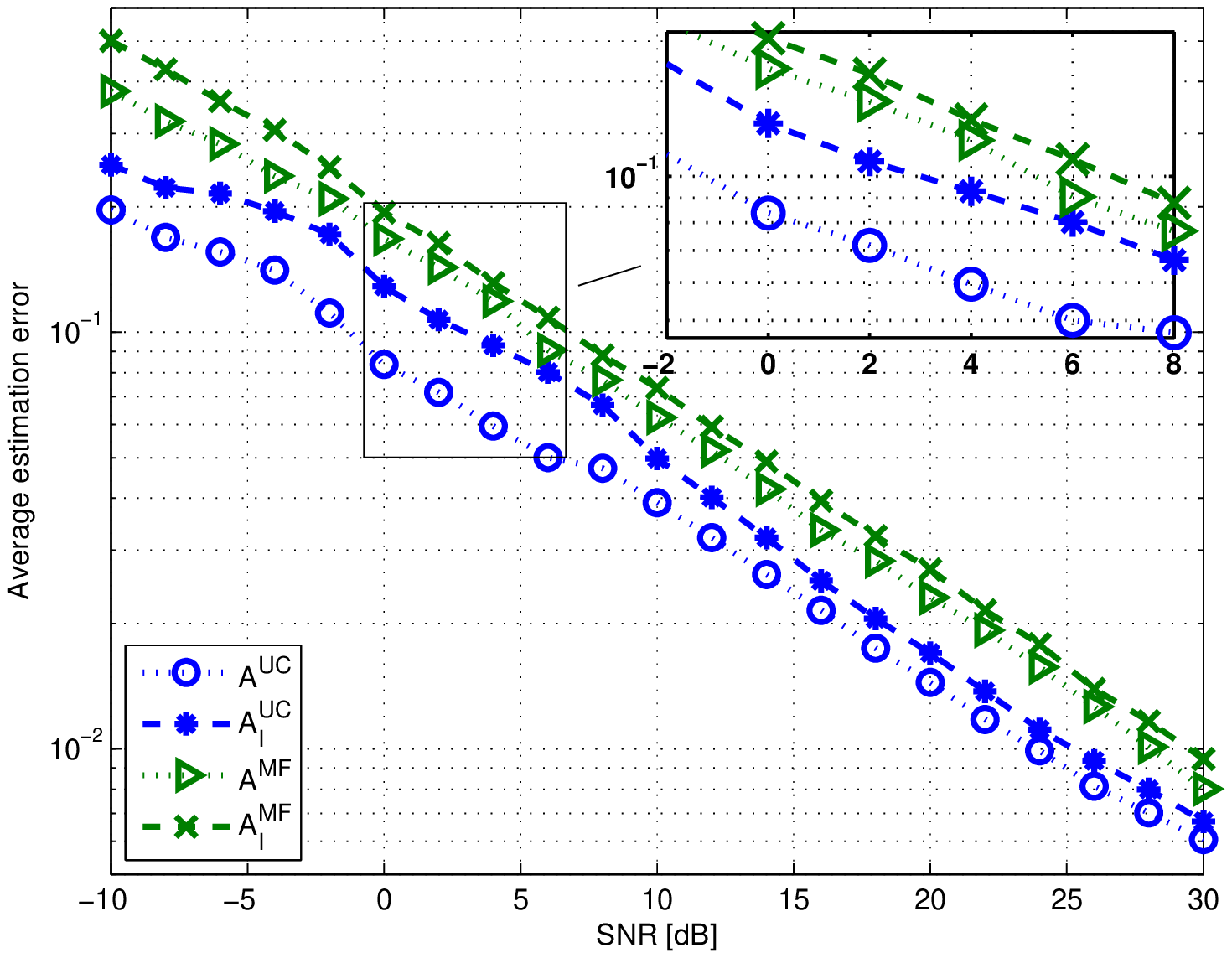}}
	\vspace{-0.6cm}
	\caption{Average estimation error vs. \ac{snr} for $\myU_S$, $\lenY = 6 \lenU$.}
	\vspace{-0.6cm}
	\label{fig:NumEx4a}
\end{figure}

Lastly, we numerically evaluate the performance gain obtained by optimizing over the unitary matrix $\myMat{V}$, detailed in Subsection \ref{subsec:AlgoSummary}. To that aim, we set ${\myA}^{\rm UC}_{\rm I}$ and ${\myA}^{\rm MF}_{\rm I}$ to be the matrices obtained via Propositions \ref{pro:KRP} and \ref{pro:KRP2}, respectively, with the unitary matrix $\myMat{V}$ fixed to $\myI_{\lenY}$. 
In Table \ref{tbl:FroComp} we detail the values of Frobenius norm $\|\myMat{V} \tilde{\myA}\subopt \! - \! \myS_{\lenY}\left(\myA \otimes \myA^* \right)\|$ computed for ${\myA}^{\rm UC}_{\rm I}$ and ${\myA}^{\rm MF}_{\rm I}$ with $\myMat{V} =\myI_{\lenY}$, and for ${\myA}^{\rm UC}$ and ${\myA}^{\rm MF}$ with $\myMat{V}$ obtained via \eqref{eqn:Procrustes1}, for $\lenY = 6 \lenU$, \ac{soi} distribution $\myU_S$, and  $1/\SigW = -10, 10, 30$ dB. 
We note that optimizing over the unitary transformation  decreases the Frobenius norm by a factor of approximately $3.3$ for  ${\myA}^{\rm UC}$ and $1.4$ for ${\myA}^{\rm MF}$.
	\begin{table}
		\centering
		\caption{Frobenius norm $\|\myMat{V} \tilde{\myA}\subopt \! - \! \myS_{\lenY}\left(\myA \otimes \myA^* \right)\|$ comparison for $\myU_S$.}
		\vspace{-0.2cm}
		\label{tbl:FroComp}
		{	
				\begin{tabular}{|c | c c c c|}
					\hline
					$1_{\vphantom{(A)}}^{\vphantom{(A)}}/\SigW $ & ${\myA}^{\rm UC}$ & ${\myA}^{\rm UC}_{\rm I}$ & ${\myA}^{\rm MF}$ & ${\myA}^{\rm MF}_{\rm I}$ \\
					\hline
					$-10$ dB & $2.09$ & $6.93$ & $6.25$ & $7.63$ \\
					$10$ dB & $2.19$ & $6.99$ & $5.70$ & $7.66$ \\
					$30$ dB & $2.25$ & $7.08$ & $5.16$ & $7.66$ \\ \hline
				\end{tabular}
			}
			\vspace{-0.2cm}
		\end{table}
To illustrate that the Frobenius norm improvement translates into improvement in estimation performance, we depict in Fig. \ref{fig:NumEx4a} the estimation error obtained with phasecut for the same setup for $1/\SigW \in [-10,30]$ dB. 
 We observe that at $\epsilon = 0.1$ optimizing the unitary matrix yields an \ac{snr} gain of $4$ dB for ${\myA}^{\rm UC}$ compared to ${\myA}^{\rm UC}_{\rm I}$, and a gain of $2$ dB for ${\myA}^{\rm MF}$ compared to ${\myA}^{\rm MF}_{\rm I}$. 
Figure \ref{fig:NumEx4a} demonstrates the benefits of optimizing over $\myMat{V}$ in Algorithm \ref{alg:Algo1} rather than choosing a fixed $\myMat{V}$.

The results of the simulation study indicate that significant performance gains can be achieved by the proposed measurement matrix design, for various recovery algorithms, using deterministic and practical measurement setups.

\vspace{-0.2cm}
\section{Conclusions}
\label{sec:Conclusions}
\vspace{-0.1cm}
In this paper we studied the design of measurement matrices for the noisy phase retrieval setup by maximizing the \ac{mi} between the \ac{soi} and the observations. Necessary conditions on the optimal measurement matrix were 
derived, and the optimal measurement matrix for Kronecker symmetric \ac{soi} in the low \ac{snr} regime was obtained in closed-form. We also studied the design of practical measurement matrices based on  maximizing the \ac{mi} between the \ac{soi} and the observations, by applying a series of approximations. 
Simulation results demonstrate the benefits of using the proposed approach for various recovery algorithms.

\vspace{-0.4cm}
\begin{appendix}
\vspace{-0.1cm}	
We first recall the definition of the Kronecker product: 
\begin{definition}[Kroncker product]
	For any $n_1 \times n_2$ matrix $\myMat{N}$ and $m_1 \times m_2$ matrix $\myMat{M}$, 	for every $p_1 \in \{1,2,\ldots,n_1\},$ $p_2 \in \{1,2,\ldots,n_2\}$, $q_1 \in \{1,2,\ldots,m_1\}$, $q_2 \in \{1,2,\ldots,m_2\}$, the entries of $\myMat{N} \otimes \myMat{M}$ are given by \cite[Ch. 1.3.6]{Golub:13}:
	\begin{equation}
	\left( \myMat{N} \otimes \myMat{M}\right)_{(p_1\!-\!1)m_1 \!+\! q_1, (p_2\!-\!1)m_2\! +\! q_2}\! =\! \left( \myMat{N}  \right)_{p_1,p_2}  \! \left( \myMat{M}  \right)_{q_1,q_2}. 
	\label{eqn:KronDef} 
	\end{equation}
\end{definition}
The following properties of the Kronecker product are repeatedly used in the sequel: 
\begin{lemma}
	\label{lem:KronIdx}
	The Kronecker product satisfies:
	\begin{enumerate}[leftmargin=0.6cm,itemindent=.0cm,labelsep=.2cm, label={\em P\arabic*}]
		\item \label{itm:A2} For any $n_1^2 \times 1$ vector $\myDetVec{x}_1$ and $n_1 \times 1$ vectors $\myDetVec{x}_2, \myDetVec{x}_3$:
		\vspace{-0.1cm}
		\begin{equation}
		\left\|\myDetVec{x}_1 - \myDetVec{x}_2 \otimes \myDetVec{x}_3^* \right\|^2 
		= \left\|{\rm vec}_{n_1}^{-1}\left( \myDetVec{x}_1\right)  - \myDetVec{x}_3^*\myDetVec{x}_2^T  \right\|^2.
			\label{eqn:KronProp1}
			\vspace{-0.1cm}
		\end{equation}
		\item \label{itm:A3} For any $\lenU \times 1$ vector $\myDetVec{x}$ and $\lenU^2 \times \lenU^2$ matrix $\myDetMat{M}$ we have that for every $k \in \lenUSet$, 
		\begin{subequations}
			\label{eqn:KronIdx}
			\vspace{-0.1cm}
\ifsingle
			\begin{equation}
			\label{eqn:KronIdx1}
			\Big( \left( {\myMat{I}_n} \otimes \myDetVec{x}^T \right)\cdot\myDetMat{M}\cdot\left( \myDetVec{x} \otimes \myDetVec{x}^* \right) \Big)_k 
			\!=\!\sum\limits_{p_1 \!=\! 1}^\lenU \sum\limits_{q_1 \!= \!1}^\lenU \sum\limits_{q_2 \!=\! 1}^\lenU \left( {\bf{x}} \right)_{q_2}\!\left( \myDetMat{M} \right)_{\left( k \! - \! 1 \right)\lenU \! + \! {q_2},\left(p_1 \! - \! 1 \right)\lenU \! + \! q_1}\! \left( \myDetVec{x} \right)_{p_1}\!  \left( \myDetVec{x}\right)^*_{q_1},
			\vspace{-0.1cm}
			\end{equation}
\else			
			\begin{align}
			\label{eqn:KronIdx1}
			&\hspace{-0.8cm}\Big( \left( {\myMat{I}_n} \otimes \myDetVec{x}^T \right)\cdot\myDetMat{M}\cdot\left( \myDetVec{x} \otimes \myDetVec{x}^* \right) \Big)_k \notag \\
			&\hspace{-0.8cm}\!=\!\sum\limits_{p_1 \!=\! 1}^\lenU \sum\limits_{q_1 \!= \!1}^\lenU \sum\limits_{q_2 \!=\! 1}^\lenU \left( {\bf{x}} \right)_{q_2}\!\left( \myDetMat{M} \right)_{\left( k \! - \! 1 \right)\lenU \! + \! {q_2},\left(p_1 \! - \! 1 \right)\lenU \! + \! q_1}\! \left( \myDetVec{x} \right)_{p_1}\!  \left( \myDetVec{x}\right)^*_{q_1},
			\vspace{-0.1cm}
			\end{align}
\fi 
		and also
		\vspace{-0.1cm}
\ifsingle
		\begin{equation}
		\label{eqn:KronIdx2}
		\Big( \left( \myDetVec{x}^T \otimes {\myMat{I}_n} \right)\cdot\myDetMat{M}^*\cdot\left( \myDetVec{x}^* \otimes \myDetVec{x} \right) \Big)_k
		\! = \! \sum\limits_{p_1 \! = \! 1}^\lenU \sum\limits_{q_1 \! = \! 1}^\lenU \sum\limits_{p_2 \! = \! 1}^\lenU \left( {\bf{x}} \right)_{p_2}\!\left( \myDetMat{M} \right)^*_{\left( p_2 \! - \! 1 \right)\lenU \! + \! k,\left( p_1 \! - \! 1 \right)\lenU \! + \! q_1}\! \left( \myDetVec{x} \right)^*_{p_1} \! \left( \myDetVec{x} \right)_{q_1}.
		\vspace{-0.1cm}
		\end{equation}	
\else		
		\begin{align}
		\label{eqn:KronIdx2}
		&\hspace{-0.8cm}\Big( \left( \myDetVec{x}^T \otimes {\myMat{I}_n} \right)\cdot\myDetMat{M}^*\cdot\left( \myDetVec{x}^* \otimes \myDetVec{x} \right) \Big)_k \notag \\
		&\hspace{-0.8cm}\! = \! \sum\limits_{p_1 \! = \! 1}^\lenU \sum\limits_{q_1 \! = \! 1}^\lenU \sum\limits_{p_2 \! = \! 1}^\lenU \left( {\bf{x}} \right)_{p_2}\!\left( \myDetMat{M} \right)^*_{\left( p_2 \! - \! 1 \right)\lenU \! + \! k,\left( p_1 \! - \! 1 \right)\lenU \! + \! q_1}\! \left( \myDetVec{x} \right)^*_{p_1} \! \left( \myDetVec{x} \right)_{q_1}.
		\vspace{-0.1cm}
		\end{align}	
\fi 
		\end{subequations}	 
	\end{enumerate}
\end{lemma}
\noindent
{\em Proof:}
Property \ref{itm:A2} follows since 
\begin{align}
\left\|\myDetVec{x}_1 - \myDetVec{x}_2 \otimes \myDetVec{x}_3^* \right\|^2 
&\stackrel{(a)}{=}  \left\|{\rm vec}_{n_1}^{-1}\left( \myDetVec{x}_1- \myDetVec{x}_2 \otimes \myDetVec{x}_3^*\right)  \right\|^2 \notag \\
&\stackrel{(b)}{=} \left\|{\rm vec}_{n_1}^{-1}\left( \myDetVec{x}_1\right)  - \myDetVec{x}_3^*\myDetVec{x}_2^T  \right\|^2,
\label{eqn:KronProp1proof}
\end{align}
where $(a)$ follows from the relationship between the Frobenious norm and the Euclidean norm, as for any square matrix $\myMat{X}$, $\|\myMat{X}\|^2 = \|{\rm vec}\left( \myMat{X}\right) \|^2$; $(b)$ follows from \cite[Ch. 12.3.4]{Golub:13}. 
		
In the proof of Property \ref{itm:A3}, we detail only the proof of \eqref{eqn:KronIdx1}, as the proof of \eqref{eqn:KronIdx2} follows using similar steps: By explicitly writing the product of the $\lenU \times \lenU^2$ matrix $ \left( {\myMat{I}_n} \otimes \myDetVec{x}^T \right)\myDetMat{M}$ and the $\lenU^2 \times 1$ vector $ \myDetVec{x} \otimes \myDetVec{x}^*$ we have that 
\ifsingle
	\begin{align}
	&\Big( \left( {\myMat{I}_n} \otimes \myDetVec{x}^T \right)\cdot\myDetMat{M}\cdot\left( \myDetVec{x} \otimes \myDetVec{x}^* \right) \Big)_k \notag \\
	&\quad= \sum\limits_{p_1 = 1}^\lenU \sum\limits_{q_1 = 1}^\lenU \left(  \left( {\myMat{I}_n} \otimes \myDetVec{x}^T \right)\cdot\myDetMat{M}\right)_{k, \left(p_1 \! - \! 1 \right)\lenU \! + \! q_1} \left( \myDetVec{x} \otimes \myDetVec{x}^*\right)_{\left(p_1 \! - \! 1 \right)\lenU \! + \! q_1}  \notag \\
	&\quad= \sum\limits_{p_1 = 1}^\lenU \sum\limits_{q_1 = 1}^\lenU \sum\limits_{p_2 = 1}^\lenU \sum\limits_{q_2 = 1}^\lenU  \left( {\myMat{I}_n} \otimes \myDetVec{x}^T \right)_{k, \left(p_2 \! - \! 1 \right)\lenU \! + \! q_2} \left(\myDetMat{M}\right)_{\left(p_2 \! - \! 1 \right)\lenU \! + \! q_2, \left(p_1 \! - \! 1 \right)\lenU \! + \! q_1} \left( \myDetVec{x} \otimes \myDetVec{x}^*\right)_{\left(p_1 \! - \! 1 \right)\lenU \! + \! q_1}. 
	\label{eqn:KronIdxProof1}
	\end{align}
\else
	\begin{align}
	&\Big( \left( {\myMat{I}_n} \otimes \myDetVec{x}^T \right)\cdot\myDetMat{M}\cdot\left( \myDetVec{x} \otimes \myDetVec{x}^* \right) \Big)_k \notag \\
	&\quad= \sum\limits_{p_1 = 1}^\lenU \sum\limits_{q_1 = 1}^\lenU \left(  \left( {\myMat{I}_n} \otimes \myDetVec{x}^T \right)\cdot\myDetMat{M}\right)_{k, \left(p_1 \! - \! 1 \right)\lenU \! + \! q_1} \left( \myDetVec{x} \otimes \myDetVec{x}^*\right)_{\left(p_1 \! - \! 1 \right)\lenU \! + \! q_1}  \notag \\
	&\quad= \sum\limits_{p_1 = 1}^\lenU \sum\limits_{q_1 = 1}^\lenU \sum\limits_{p_2 = 1}^\lenU \sum\limits_{q_2 = 1}^\lenU  \left( {\myMat{I}_n} \otimes \myDetVec{x}^T \right)_{k, \left(p_2 \! - \! 1 \right)\lenU \! + \! q_2} \notag \\
	&\qquad\quad  \cdot \left(\myDetMat{M}\right)_{\left(p_2 \! - \! 1 \right)\lenU \! + \! q_2, \left(p_1 \! - \! 1 \right)\lenU \! + \! q_1} \left( \myDetVec{x} \otimes \myDetVec{x}^*\right)_{\left(p_1 \! - \! 1 \right)\lenU \! + \! q_1}. 
	\label{eqn:KronIdxProof1}
	\end{align}
\fi 
Next, from \eqref{eqn:KronDef}  we have that $\left( {\myMat{I}_n} \otimes \myDetVec{x}^T \right)_{k, \left(p_2 \! - \! 1 \right)\lenU \! + \! q_2} \!=\! \left(  {\myMat{I}_n}\right)_{k,p_2} \cdot \left(  \myDetVec{x}\right) _{q_2} \!=\! \delta_{k,p_2} \left(  \myDetVec{x}\right) _{q_2}$ and  $\left( \myDetVec{x} \otimes \myDetVec{x}^*\right)_{\left(p_1 \! - \! 1 \right)\lenU \! + \! q_1}\! =\! \left( \myDetVec{x}\right) _{p_1}\cdot\left( \myDetVec{x}\right)^*_{q_1}$. Substituting these computations back into \eqref{eqn:KronIdxProof1} yields 
\ifsingle
	\begin{equation*}
	\Big( \left( {\myMat{I}_n} \otimes \myDetVec{x}^T \right)\cdot\myDetMat{M}\cdot\left( \myDetVec{x} \otimes \myDetVec{x}^* \right) \Big)_k 
	=  \sum\limits_{p_1 \!=\! 1}^\lenU \sum\limits_{q_1 \!= \!1}^\lenU \sum\limits_{q_2 \!=\! 1}^\lenU \left( {\bf{x}} \right)_{q_2}\!\left( \myDetMat{M} \right)_{\left( k \! - \! 1 \right)\lenU \! + \! {q_2},\left(p_1 \! - \! 1 \right)\lenU \! + \! q_1}\! \left( \myDetVec{x} \right)_{p_1}\!  \left( \myDetVec{x}\right)^*_{q_1},
	\end{equation*}
\else
	\begin{align*}
	&\Big( \left( {\myMat{I}_n} \otimes \myDetVec{x}^T \right)\cdot\myDetMat{M}\cdot\left( \myDetVec{x} \otimes \myDetVec{x}^* \right) \Big)_k \notag \\
	&\quad=  \sum\limits_{p_1 \!=\! 1}^\lenU \sum\limits_{q_1 \!= \!1}^\lenU \sum\limits_{q_2 \!=\! 1}^\lenU \left( {\bf{x}} \right)_{q_2}\!\left( \myDetMat{M} \right)_{\left( k \! - \! 1 \right)\lenU \! + \! {q_2},\left(p_1 \! - \! 1 \right)\lenU \! + \! q_1}\! \left( \myDetVec{x} \right)_{p_1}\!  \left( \myDetVec{x}\right)^*_{q_1},
	\end{align*}
\fi 
proving \eqref{eqn:KronIdx1}. 
\qed
\vspace{-0.15cm}
\subsection{Proof of Theorem \ref{thm:GeneralSource}}
\label{app:GeneralSource}
\vspace{-0.1cm}
Applying the KKT theorem \cite[Ch. 5.5.3]{Boyd:04} to the problem \eqref{eqn:MIProbDef1}, we obtain the following necessary conditions for  $\myA\opt$:
\begin{subequations}
	\label{eqn:KKTcond}
	\begin{equation}
	\label{eqn:KKTcond1}
\hspace{-0.2cm}	\left. {\nabla _{\myA}}\Big( - I\left( \myU;\myY \right) \!-\! \lambda \left( \myP\! -\! {\rm Tr}\left( \myA\myA^H \right) \right) \Big) \right|_{ \myA= \myA\opt} \!= 0,
	\end{equation}
	and
	\begin{equation}
	\label{eqn:KKTcond2}
	\lambda \Big( {\myP - {\rm Tr}\left(\myA\opt\left( \myA\opt\right) ^H \right)} \Big) = 0,
	\end{equation}
\end{subequations}
where $\lambda \geq 0$. From \eqref{eqn:KKTcond1} it follows that for $\myA = \myA\opt$
\begin{align}
\left. {\nabla _{\myA}}\Big( I\left( \myU;\myY \right)\Big)  \right|_{ \myA= \myA\opt}
&= \left. \lambda\cdot {\nabla _{\myA}}\Big( {\rm Tr}\left( \myA\myA^H \right)\Big) \right|_{ \myA= \myA\opt} \notag \\
&= \lambda \cdot\myA\opt.
\label{eqn:KKTDer1}
\end{align}

To determine the derivative of the left-hand side of \eqref{eqn:KKTDer1}, we use the chain rule for complex gradients \cite[Ch. 4.1.1]{Petersen:08}, from which we have that for every $k_1 \!\in\! \lenYSet, k_2 \!\in\! \lenUSet$, 
\begin{align}
&\hspace{-0.2cm}\left( {\nabla _{\myA}}\Big( I\left( \myU;\myY \right)\Big) \right)_{k_1,k_2}
\!\!\!\!=\! {\rm Tr}\left(\left({\nabla _{\tilde{\myA}}}\Big( I\left( \myU;\myY \right)\Big) \right)^T\! \frac{\partial \tilde{\myA}^*}{\partial \left(\myA \right)_{k_1,k_2}^* }  \right) \notag \\
&\qquad\qquad  + {\rm Tr}\left(\left({\nabla _{\tilde{\myA}^*}}\Big( I\left( \myU;\myY \right)\Big) \right)^T \!\frac{\partial \tilde{\myA}}{\partial \left(\myA \right)_{k_1,k_2}^* }  \right).
\label{eqn:KKTDer2a}
\end{align}
Next, we let $\Emat_C\left(\myA \right)$ denote the  \ac{mmse} matrix for estimating $\myKron$ from $\myY_C$, 
and note that \eqref{eqn:MIEq1} implies that
\begin{align}
{\nabla _{\tilde{\myA}}}\Big( I\left( \myU;\myY \right)\Big)  
&= {\nabla _{\tilde{\myA}}}\Big( I\left( \myKron;\myY_C \right)\Big)  \notag \\
&\stackrel{(a)}{=} \tilde{\myA} \cdot \Emat_C\left(\myA \right) \stackrel{(b)}{=} \tilde{\myA} \cdot \Emat\left(\myA \right) ,
\label{eqn:KKTDer1a}
\end{align}
where $(a)$ follows from \cite[Eq. (4)]{Palomar:06}, since the relationship between $\myY_C$ and $\myKron$ corresponds to a  \ac{pc} Gaussian \ac{mimo} channel with input $\myKron$ and output $\myY_C$; 
$(b)$ follows since $\myW_I \! = \! {\rm Im}\left\{\myY_C \right\}$ is independent of $\myY \! = \! {\rm Re}\left\{\myY_C \right\}$ and of $\myKron$, thus the \ac{mmse} matrix for estimating $\myKron$ from $\myY_C$,  $\Emat_C\left(\myA\right)$, is equal to the \ac{mmse} matrix for estimating $\myKron$ from $\myY$, $\Emat\left(\myA \right)$.
As \ac{mi} is real-valued, it follows from \eqref{eqn:KKTDer1a} and from the definition of the generalized complex derivative \cite[Ch. 4.1.1]{Petersen:08} that 
\begin{equation}
{\nabla _{\tilde{\myA}^*}}\Big( I\left( \myU;\myY \right)\Big)  
= \left( \tilde{\myA} \cdot \Emat\left(\myA \right)\right)^*.
\label{eqn:KKTDer1b}
\end{equation} 
Plugging \eqref{eqn:KKTDer1a} and \eqref{eqn:KKTDer1b} into \eqref{eqn:KKTDer2a} results in
\begin{align}
&\hspace{-0.4cm}\left( {\nabla _{\myA}}\Big( I\left( \myU;\myY \right)\Big) \right)_{k_1,k_2}
\!=\! \sum\limits_{{l_1} = 1}^\lenY \sum\limits_{{l_2} = 1}^{\lenU^2} \left( \tilde{\myA} \cdot\Emat\left(\myA \right)  \right)_{{l_1},{l_2}}\frac{\partial \left( \tilde{\myA} \right)_{{l_1},{l_2}}^*}{\partial \left( \myA \right)_{{k_1},{k_2}}^*}\notag \\
&\qquad \qquad\quad + \sum\limits_{{l_1} = 1}^\lenY \sum\limits_{{l_2} = 1}^{\lenU^2} \left( \tilde{\myA}\cdot\Emat\left(\myA \right)  \right)^*_{{l_1},{l_2}}\frac{\partial \left( \tilde{\myA} \right)_{{l_1},{l_2}}}{\partial \left( \myA \right)_{{k_1},{k_2}}^*}.
\label{eqn:KKTDer2}
\end{align}
%
By writing the index $l_2$ as $l_2\! =\! (p_2\!-\!1) \lenU\! +\! q_2$, where $p_2,q_2 \!\in\! \lenUSet$, it follows from the definition of $\tilde{\myA}$ in \eqref{eqn:AtildeDef} that 
\begin{subequations}
\label{eqn:PartialDer}
\begin{equation}	
\label{eqn:PartialDera}
\frac{\partial \big( \tilde{\myA} \big)_{{l_1},(p_2\! - \!1) \lenU \! + \! q_2}^*}{\partial \left( \myA \right)_{{k_1},{k_2}}^*} \! = \! \left( \myA \right)_{{k_1},q_2} \delta_{l_1,k_1}\delta_{p_2,k_2}, 
\end{equation}
and  
\begin{equation}	
\label{eqn:PartialDerb}
\frac{\partial \big( \tilde{\myA} \big)_{{l_1},(p_2\! - \!1) \lenU \! + \! q_2}}{\partial \left( \myA \right)_{{k_1},{k_2}}^*} \! = \! \left( \myA \right)_{{k_1},p_2} \delta_{l_1,k_1}\delta_{q_2,k_2}. 
\end{equation}
\end{subequations}
Thus, \eqref{eqn:KKTDer2} yields
\vspace{-0.2cm}
\begin{align}
&\hspace{-0.4cm}\left( {\nabla _{\myA}}\Big( I\left( \myU;\myY \right)\Big) \right)_{k_1,k_2}  
\!=\! \sum\limits_{q_2 = 1}^{\lenU} \left( \tilde{\myA} \cdot\Emat\left(\myA \right)  \right)_{{k_1},(k_2\!-\!1)\lenU \!+\! q_2}\!\!\left( \myA \right)_{{k_1},q_2}  \notag \\
&\qquad \qquad \quad+ \sum\limits_{p_2 = 1}^{\lenU} \left( \tilde{\myA}\cdot \Emat\left(\myA \right)  \right)_{{k_1},(p_2\!-\!1)\lenU\! +\! k_2}^*\!\!\left( \myA \right)_{{k_1},p_2}.
\label{eqn:KKTDer3}
\vspace{-0.2cm}
\end{align}

Next, we note that 
\ifsingle 
	\begin{align}
	\left( \tilde{\myA} \cdot\Emat\left(\myA \right)  \right)_{{k_1},(p_2\! - \!1)\lenU \! + \! q_2}
	&\! = \!  \sum\limits_{{p_1} \! = \! 1}^\lenU \sum\limits_{{q_1} \! = \! 1}^\lenU \big( \tilde{\myA} \big)_{{k_1},\left( {{p_1} \! - \! 1} \right)\lenU \! + \! {q_1}}\big( \Emat\left(\myA \right)  \big)_{\left( {{p_1} \! - \! 1} \right)\lenU \! + \! {q_1},\left( {{p_2} \! - \! 1} \right)\lenU \! + \! {q_2}}  \notag \\
	&\hspace{-0.4cm} \!\stackrel{(a)}{ = }\!  \sum\limits_{{p_1} \! = \! 1}^\lenU \sum\limits_{{q_1} \! = \! 1}^\lenU \!{{\left( \myA \right)}_{{k_1},{p_1}}}\!\left( \myA \right)_{{k_1},{q_1}}^*\!\big( \Emat\left(\myA \right)  \big)_{\left( {{p_1} \! - \! 1} \right)\lenU \! + \! {q_1},\left( {{p_2} \! - \! 1} \right)\lenU \! + \! {q_2}}, 
	\label{eqn:KKTDer4}  
	\end{align}
\else
	\begin{align}
	&\hspace{-0.4cm}\left( \tilde{\myA} \cdot\Emat\left(\myA \right)  \right)_{{k_1},(p_2\! - \!1)\lenU \! + \! q_2} \notag \\
	&\hspace{-0.4cm}\! = \!  \sum\limits_{{p_1} \! = \! 1}^\lenU \sum\limits_{{q_1} \! = \! 1}^\lenU \big( \tilde{\myA} \big)_{{k_1},\left( {{p_1} \! - \! 1} \right)\lenU \! + \! {q_1}}\big( \Emat\left(\myA \right)  \big)_{\left( {{p_1} \! - \! 1} \right)\lenU \! + \! {q_1},\left( {{p_2} \! - \! 1} \right)\lenU \! + \! {q_2}}  \notag \\
	&\hspace{-0.4cm} \!\stackrel{(a)}{ = }\!  \sum\limits_{{p_1} \! = \! 1}^\lenU \sum\limits_{{q_1} \! = \! 1}^\lenU \!{{\left( \myA \right)}_{{k_1},{p_1}}}\!\left( \myA \right)_{{k_1},{q_1}}^*\!\big( \Emat\left(\myA \right)  \big)_{\left( {{p_1} \! - \! 1} \right)\lenU \! + \! {q_1},\left( {{p_2} \! - \! 1} \right)\lenU \! + \! {q_2}}, 
	\label{eqn:KKTDer4}  
	\end{align}
\fi 
where $(a)$ follows from the definition of $\tilde{\myA}$ in \eqref{eqn:AtildeDef}.
Plugging \eqref{eqn:KKTDer3} and \eqref{eqn:KKTDer4} into \eqref{eqn:KKTDer1}, we conclude that the entries of the optimal measurement matrix $\myA\opt$ satisfy 
\ifsingle 
	\begin{align}
	\lambda\cdot  \left(\myA\opt \right)_{k_1,k_2}\!\!
	&\! = \!  \sum\limits_{q_2 \! = \! 1}^{\lenU}  \sum\limits_{{p_1} \! = \! 1}^\lenU \sum\limits_{{q_1} \! = \! 1}^\lenU \!{\left( \myA\opt \right)}_{{k_1},{p_1}}\!\left( \myA\opt \right)_{{k_1},{q_1}}^*\!\left( \myA\opt \right)_{{k_1},q_2}
	 \Big( \Emat\left(\myA\opt \right)  \Big)_{\left( {{p_1} \! - \! 1} \right)\lenU \! + \! {q_1},\left( {{k_2} \! - \! 1} \right)\lenU \! + \! {q_2}} \notag \\
	&\! + \! \sum\limits_{p_2 \! = \! 1}^{\lenU}  \sum\limits_{{p_1} \! = \! 1}^\lenU \sum\limits_{{q_1} \! = \! 1}^\lenU \left( \myA\opt \right)_{{k_1},{p_1}}^*\!\left( \myA\opt \right)_{{k_1},{q_1}}\!\left( \myA\opt \right)_{{k_1},p_2}
	 \Big( \Emat\left(\myA\opt \right)  \Big)_{\left( {{p_1} \! - \! 1} \right)\lenU \! + \! {q_1},\left( {{p_2} \! - \! 1} \right)\lenU \! + \! {k_2}}^*,
	\label{eqn:KKTDer5}  
	\end{align}
\else
	\begin{align}
	\lambda\cdot  \left(\myA\opt \right)_{k_1,k_2}\!\!
	&\! = \!  \sum\limits_{q_2 \! = \! 1}^{\lenU}  \sum\limits_{{p_1} \! = \! 1}^\lenU \sum\limits_{{q_1} \! = \! 1}^\lenU \!{\left( \myA\opt \right)}_{{k_1},{p_1}}\!\left( \myA\opt \right)_{{k_1},{q_1}}^*\!\left( \myA\opt \right)_{{k_1},q_2}\notag \\
	&\qquad\quad \cdot \Big( \Emat\left(\myA\opt \right)  \Big)_{\left( {{p_1} \! - \! 1} \right)\lenU \! + \! {q_1},\left( {{k_2} \! - \! 1} \right)\lenU \! + \! {q_2}} \notag \\
	&\! + \! \sum\limits_{p_2 \! = \! 1}^{\lenU}  \sum\limits_{{p_1} \! = \! 1}^\lenU \sum\limits_{{q_1} \! = \! 1}^\lenU \left( \myA\opt \right)_{{k_1},{p_1}}^*\!\left( \myA\opt \right)_{{k_1},{q_1}}\!\left( \myA\opt \right)_{{k_1},p_2}\notag \\
	&\qquad \quad\cdot \Big( \Emat\left(\myA\opt \right)  \Big)_{\left( {{p_1} \! - \! 1} \right)\lenU \! + \! {q_1},\left( {{p_2} \! - \! 1} \right)\lenU \! + \! {k_2}}^*,
	\label{eqn:KKTDer5}  
	\end{align}
\fi 
where $\lambda$ is set to satisfy the power constraint.

We now use Property \ref{itm:A3} of Lemma \ref{lem:KronIdx} to express \eqref{eqn:KKTDer5} in vector form. Letting $\Avec\opt_k$ denote the $k$-th column of $\left( \myA\opt\right)^T$, we note that the first and second summands in the right hand side of \eqref{eqn:KKTDer5} correspond to \eqref{eqn:KronIdx1} and \eqref{eqn:KronIdx2}, respectively, with $\myDetVec{x} = \Avec\opt_{k_1}$ and $\myMat{M} = \Emat^T\!\left(\myA\opt \right)$. Thus,  \eqref{eqn:KKTDer5}  can be written as 
\ifsingle 
	\begin{align}
	\lambda \cdot\left(\myA\opt \right)_{k_1,k_2}  
	& = \bigg(\! \left( \myI_n \otimes \left( \Avec\opt_{k_1}\right) ^T\right)\!\cdot \!\Emat^T\!\left(\myA\opt \right)\!\cdot \!\left(\Avec\opt_{k_1} \otimes \left( \Avec\opt_{k_1}\right) ^* \right)\!\bigg)_{k_2} \notag \\
	&\quad+ \bigg(\! \left( \left( \Avec\opt_{k_1}\right) ^T\! \otimes \myI_n \right)\! \cdot\!\Emat^H\!\left(\myA\opt \right) \!\cdot\!\left(\left( \Avec\opt_{k_1}\right) ^* \!\otimes  \Avec\opt_{k_1} \right)\!\bigg)_{k_2}. 
	\label{eqn:KKTDer6}  
	\end{align}
\else
	\begin{align}
	&\hspace{-0.2cm}\lambda \cdot\left(\myA\opt \right)_{k_1,k_2}  \notag \\
	&\hspace{-0.2cm} = \bigg(\! \left( \myI_n \otimes \left( \Avec\opt_{k_1}\right) ^T\right)\!\cdot \!\Emat^T\!\left(\myA\opt \right)\!\cdot \!\left(\Avec\opt_{k_1} \otimes \left( \Avec\opt_{k_1}\right) ^* \right)\!\bigg)_{k_2} \notag \\
	&+ \bigg(\! \left( \left( \Avec\opt_{k_1}\right) ^T\! \otimes \myI_n \right)\! \cdot\!\Emat^H\!\left(\myA\opt \right) \!\cdot\!\left(\left( \Avec\opt_{k_1}\right) ^* \!\otimes  \Avec\opt_{k_1} \right)\!\bigg)_{k_2}. 
	\label{eqn:KKTDer6}  
	\end{align}
\fi 
Consequently, as the \ac{mmse} matrix is Hermitian, we have 
\ifsingle 
	\begin{align}
	\lambda\cdot\Avec\opt_{k_1} \!
	&= \left( \myI_n \otimes \left( \Avec\opt_{k_1}\right) ^T\right)\cdot \Emat^T\!\left(\myA\opt \right)\cdot \left(\Avec\opt_{k_1} \otimes \left( \Avec\opt_{k_1}\right) ^* \right)+ \left( \left( \Avec\opt_{k_1}\right) ^T \otimes \myI_n \right)\cdot \Emat\!\left(\myA\opt \right)\cdot \left( \left( \Avec\opt_{k_1}\right) ^* \otimes \Avec\opt_{k_1} \right) \notag \\
	&= \bigg(\! \left( \myI_n\! \otimes \!\left( \Avec\opt_{k_1}\right)^T\right)\cdot \Emat^T\!\left(\myA\opt \right)\cdot \left(\myI_n \! \otimes \left( \Avec\opt_{k_1}\right)^* \right) 
	\!+\! \left( \left( \Avec\opt_{k_1}\right)^T\! \otimes\! \myI_n \right)\cdot \Emat\left(\myA\opt \right) \cdot\left( \left( \Avec\opt_{k_1}\right)^*\! \otimes\! \myI_n  \right)\! \bigg) \Avec\opt_{k_1} \notag \\
	& =  \myDetMat{H}_{k_1}\!\left(\myA\opt \right)\cdot \Avec\opt_{k_1}, \qquad  k_1 \in \lenYSet,
	\label{eqn:KKTDer7}  
	\end{align}
\else
	\begin{align}
	\lambda\cdot\Avec\opt_{k_1} \!
	&= \left( \myI_n \otimes \left( \Avec\opt_{k_1}\right) ^T\right)\cdot \Emat^T\!\left(\myA\opt \right)\cdot \left(\Avec\opt_{k_1} \otimes \left( \Avec\opt_{k_1}\right) ^* \right) \notag \\
	&\quad + \left( \left( \Avec\opt_{k_1}\right) ^T \otimes \myI_n \right)\cdot \Emat\!\left(\myA\opt \right)\cdot \left( \left( \Avec\opt_{k_1}\right) ^* \otimes \Avec\opt_{k_1} \right) \notag \\
	&= \bigg(\! \left( \myI_n\! \otimes \!\left( \Avec\opt_{k_1}\right)^T\right)\cdot \Emat^T\!\left(\myA\opt \right)\cdot \left(\myI_n \! \otimes \left( \Avec\opt_{k_1}\right)^* \right) \notag \\
	&\quad 
	\!+\! \left( \left( \Avec\opt_{k_1}\right)^T\! \otimes\! \myI_n \right)\cdot \Emat\left(\myA\opt \right) \cdot\left( \left( \Avec\opt_{k_1}\right)^*\! \otimes\! \myI_n  \right)\! \bigg) \Avec\opt_{k_1} \notag \\
	& =  \myDetMat{H}_{k_1}\!\left(\myA\opt \right)\cdot \Avec\opt_{k_1}, \qquad  k_1 \in \lenYSet,
	\label{eqn:KKTDer7}  
	\end{align}
\fi 
proving the theorem. 
\qed

\vspace{-0.15cm}
\subsection{Proof of Lemma \ref{lem:Cond2}}
\label{app:Cond2}
We first write the indexes $k_1, k_2 \in \{1,2,\ldots,\lenU^2\}$ as $k_1 = (p_1 -1)\lenU + q_1$ and $k_2 = (p_2 -1)\lenU + q_2$, where $p_1, p_2, q_1, q_2 \in \lenUSet$. Using \eqref{eqn:KronDef}, the entries of the covariance matrix of ${\myX \otimes \myX^*}$, denoted $\CovMat{\myX \otimes \myX^*}$, can then be written as
\ifsingle 
	 \begin{align}
	 &\left( \CovMat{\myX \otimes \myX^*}\right)_{(p_1 -1)\lenU + q_1,(p_2 -1)\lenU + q_2} \notag \\
	 &\qquad= \E\Big\{\left(\myX \right)_{p_1} \left(\myX \right)_{q_1}^* \left(\myX \right)_{p_2}^* \left(\myX \right)_{q_2}\Big\} 
	 - \E\Big\{\left(\myX \right)_{p_1} \left(\myX \right)_{q_1}^* \Big\}\E\Big\{\left(\myX \right)_{p_2}^* \left(\myX \right)_{q_2} \Big\} \notag \\
	 &\qquad\stackrel{(a)}{=} \E\Big\{\left(\myX \right)_{p_1} \left(\myX \right)_{q_1}^* \Big\}\E\Big\{\left(\myX \right)_{p_2}^* \left(\myX \right)_{q_2} \Big\} 
	 + \E\Big\{\left(\myX \right)_{p_1} \left(\myX \right)_{p_2}^* \Big\}\E\Big\{\left(\myX \right)_{q_1}^* \left(\myX \right)_{q_2} \Big\} \notag \\
	 &\qquad\qquad + \E\Big\{\left(\myX \right)_{p_1} \left(\myX \right)_{q_2} \Big\}\E\Big\{\left(\myX \right)_{p_2}^* \left(\myX \right)_{p_1}^* \Big\}
	 - \E\Big\{\left(\myX \right)_{p_1} \left(\myX \right)_{q_1}^* \Big\}\E\Big\{\left(\myX \right)_{p_2}^* \left(\myX \right)_{q_2} \Big\} \notag \\
	 &\qquad\stackrel{(b)}{=}  \E\Big\{\left(\myX \right)_{p_1} \left(\myX \right)_{p_2}^* \Big\}\E\Big\{\left(\myX \right)_{q_1}^* \left(\myX \right)_{q_2} \Big\} \notag \\
	 &\qquad= \left( \CovMat{\myX}\right)_{p_1,p_2} \left( \CovMat{\myX}\right)_{q_1,q_2}^* \notag \\
	 &\qquad\stackrel{(c)}{=} \left( \CovMat{\myX} \otimes \CovMat{\myX}^*\right)_{(p_1 -1)\lenU + q_1,(p_2 -1)\lenU + q_2},
	 \label{eqn:lemCond2}
	 \end{align}
\else
	 \begin{align}
	 &\left( \CovMat{\myX \otimes \myX^*}\right)_{(p_1 -1)\lenU + q_1,(p_2 -1)\lenU + q_2} \notag \\
	 &\qquad= \E\Big\{\left(\myX \right)_{p_1} \left(\myX \right)_{q_1}^* \left(\myX \right)_{p_2}^* \left(\myX \right)_{q_2}\Big\} \notag \\
	 &\qquad\qquad 
	 - \E\Big\{\left(\myX \right)_{p_1} \left(\myX \right)_{q_1}^* \Big\}\E\Big\{\left(\myX \right)_{p_2}^* \left(\myX \right)_{q_2} \Big\} \notag \\
	 &\qquad\stackrel{(a)}{=} \E\Big\{\left(\myX \right)_{p_1} \left(\myX \right)_{q_1}^* \Big\}\E\Big\{\left(\myX \right)_{p_2}^* \left(\myX \right)_{q_2} \Big\} \notag \\
	 &\qquad\qquad 
	 + \E\Big\{\left(\myX \right)_{p_1} \left(\myX \right)_{p_2}^* \Big\}\E\Big\{\left(\myX \right)_{q_1}^* \left(\myX \right)_{q_2} \Big\} \notag \\
	 &\qquad\qquad + \E\Big\{\left(\myX \right)_{p_1} \left(\myX \right)_{q_2} \Big\}\E\Big\{\left(\myX \right)_{p_2}^* \left(\myX \right)_{p_1}^* \Big\}\notag \\
	 &\qquad\qquad  - \E\Big\{\left(\myX \right)_{p_1} \left(\myX \right)_{q_1}^* \Big\}\E\Big\{\left(\myX \right)_{p_2}^* \left(\myX \right)_{q_2} \Big\} \notag \\
	 &\qquad\stackrel{(b)}{=}  \E\Big\{\left(\myX \right)_{p_1} \left(\myX \right)_{p_2}^* \Big\}\E\Big\{\left(\myX \right)_{q_1}^* \left(\myX \right)_{q_2} \Big\} \notag \\
	 &\qquad= \left( \CovMat{\myX}\right)_{p_1,p_2} \left( \CovMat{\myX}\right)_{q_1,q_2}^* \notag \\
	 &\qquad\stackrel{(c)}{=} \left( \CovMat{\myX} \otimes \CovMat{\myX}^*\right)_{(p_1 -1)\lenU + q_1,(p_2 -1)\lenU + q_2},
	 \label{eqn:lemCond2}
	 \end{align}
\fi 
 where $(a)$ follows from Isserlis theorem for complex Gaussian random vectors \cite[Ch. 1.4]{Koopmans:93}; 
 $(b)$ follows from the proper complexity of $\myX$, which implies that $\E\big\{\left(\myX \right)_{p_1} \left(\myX \right)_{q_2} \big\}\E\big\{\left(\myX \right)_{p_2}^* \left(\myX \right)_{p_1}^* \big\}=0$; 
 and $(c)$ follows from \eqref{eqn:KronDef}.  Eq. \eqref{eqn:lemCond2} proves the lemma. 
 \qed

\vspace{-0.15cm}
\subsection{Proof of Theorem \ref{thm:GaussianSource}}
\label{app:GaussianSource}
To solve the optimization problem \eqref{eqn:MIProbDef2}, we employ the following auxiliary lemma:
\begin{lemma}
	\label{cor:Cond3} 
	Let $\Avec_k$ be the $k$-th column of $\myA^T$, $k \in \lenYSet$. 
	If $\myU$ is Kronecker symmetric  with covariance matrix $\CovMat{\myU}$, then
	\begin{equation}
	\label{eqn:Cond3}
	{\rm Tr}\left(\tilde{\myA}\CovMat{\myKron} \tilde{\myA}^H \right) = \sum\limits_{k=1}^{\lenY}\left(\Avec_k^H  \CovMat{\myU}^* \Avec_k\right)^2. 
	\end{equation}	
\end{lemma}

\noindent
{\em Proof:}
Using Def. \ref{def:KronSym} and the representation \eqref{eqn:AtildeDef2} it follows that 
\ifsingle 
	\begin{align}
	{\rm Tr}\left(\tilde{\myA}\CovMat{\myKron} \tilde{\myA}^H \right)
	&= {\rm Tr}\Big(\myS_{\lenY} \left( \myA \otimes \myA^* \right)\cdot\left( \CovMat{\myU} \otimes \CovMat{\myU}^*\right)\cdot \left( \myA^H \otimes \myA^T \right)\myS_{\lenY}^H  \Big) \notag \\
	&\stackrel{(a)}{=}  {\rm Tr}\left(\myS_{\lenY}^H  \myS_{\lenY} \left(  \left( \myA\CovMat{\myU}\myA^H\right) \otimes \left( \myA\CovMat{\myU}\myA^H\right)^* \right) \right),
	\label{eqn:Proof3_a}
	\end{align}
\else
	\begin{align}
	&{\rm Tr}\left(\tilde{\myA}\CovMat{\myKron} \tilde{\myA}^H \right) \notag \\
	&\quad= {\rm Tr}\Big(\myS_{\lenY} \left( \myA \otimes \myA^* \right)\cdot\left( \CovMat{\myU} \otimes \CovMat{\myU}^*\right)\cdot \left( \myA^H \otimes \myA^T \right)\myS_{\lenY}^H  \Big) \notag \\
	&\quad\stackrel{(a)}{=}  {\rm Tr}\left(\myS_{\lenY}^H  \myS_{\lenY} \left(  \left( \myA\CovMat{\myU}\myA^H\right) \otimes \left( \myA\CovMat{\myU}\myA^H\right)^* \right) \right),
	\label{eqn:Proof3_a}
	\end{align}
\fi 
where $(a)$ follows from the properties of the trace operator \cite[Ch, 1.1]{Petersen:08} and the Kronecker product \cite[Ch, 10.2]{Petersen:08}. 
Note that $\myS_{\lenY}^H  \myS_{\lenY} $ is an $\lenY^2 \times \lenY^2$ diagonal matrix which satisfies $\left( \myS_{\lenY}^H  \myS_{\lenY}\right)_{l,l} = 1$ if $l = (k-1)\lenY + k$ for some $k \in \lenYSet$ and $\left( \myS_{\lenY}^H  \myS_{\lenY}\right)_{l,l} = 0$ otherwise. Therefore, \eqref{eqn:Proof3_a} can be written as 
\ifsingle
	\begin{align}
	{\rm Tr}\left(\!\tilde{\myA}\CovMat{\myKron} \tilde{\myA}^H \!\right) 
	&\!=\! \sum\limits_{k=1}^{\lenY} \Big( \left( \myA\CovMat{\myU}\myA^H\right) \otimes \left( \myA\CovMat{\myU}\myA^H\right)^* \Big)_{(k-1)\lenY + k,(k-1)\lenY + k} \notag \\
	&\!\stackrel{(a)}{=}\! \sum\limits_{k=1}^{\lenY} \left| \Avec_k^T\CovMat{\myU}\Avec_k^*\right|^2 
	\!\stackrel{(b)}{=}\! \sum\limits_{k=1}^{\lenY} \left( \Avec_k^H\CovMat{\myU}^*\Avec_k\right)^2, 
	\label{eqn:Proof3_b}
	\end{align}
\else
	\begin{align}
	{\rm Tr}\left(\!\tilde{\myA}\CovMat{\myKron} \tilde{\myA}^H \!\right) 
	&\!=\! \sum\limits_{k=1}^{\lenY} \Big( \left( \myA\CovMat{\myU}\myA^H\right) \notag \\
	&\qquad \qquad  \otimes \left( \myA\CovMat{\myU}\myA^H\right)^* \Big)_{(k-1)\lenY + k,(k-1)\lenY + k} \notag \\
	&\!\stackrel{(a)}{=}\! \sum\limits_{k=1}^{\lenY} \left| \Avec_k^T\CovMat{\myU}\Avec_k^*\right|^2 
	\!\stackrel{(b)}{=}\! \sum\limits_{k=1}^{\lenY} \left( \Avec_k^H\CovMat{\myU}^*\Avec_k\right)^2, 
	\label{eqn:Proof3_b}
	\end{align}
\fi 
where $(a)$ follows from \eqref{eqn:KronDef} and from the definition of $\Avec_k$ as the $k$-th column of $\myA^T$, and $(b)$ follows since $\CovMat{\myU}$ is Hermitian and positive semi-definite.
\qed

\smallskip
Using Lemma \ref{cor:Cond3}, \eqref{eqn:MIProbDef2} can  be written as 
\begin{align}
\myA\opt 
&=\left[\Avec_1\opt, \Avec_2\opt, \ldots, \Avec_\lenY\opt\right]^T \notag \\
&= \mathop {{\rm{argmax}}}\limits_{\{\Avec_k\}_{k=1}^{\lenY}:{\rm{ }}\sum\limits_{k=1}^{\lenY} \|\Avec_k\|^2 \le \myP} \sum\limits_{k=1}^{\lenY} \left( \Avec_k^H\CovMat{\myU}^*\Avec_k\right)^2 \notag \\
&=\mathop {{\rm{argmax}}}\limits_{\{\Avec_k\}_{k=1}^{\lenY}:{\rm{ }}\sum\limits_{k=1}^{\lenY} \|\Avec_k\|^2 \le \myP} \sum\limits_{k=1}^{\lenY} \left( \frac{ \Avec_k^H \CovMat{\myU}^* \Avec_k} {\|\Avec_k\|}\right)^2\|\Avec_k\|^2.
\label{eqn:MIProbDef3}
\end{align}
The maximal value of the ratio $\frac{ \Avec_k^H \CovMat{\myU}^* \Avec_k} {\|\Avec_k\|}$ is the largest eigenvalue of  $\CovMat{\myU}^*$, denoted $\mu_{\max}$.
This maximum is obtained by setting  $\frac{ \Avec_k} {\|\Avec_k\|} = e^{j2\pi\phi_k} \myVec{v}_{\max}^*$, where $\myVec{v}_{\max}^*$ is the eigenvector of $\CovMat{\myU}^*$ corresponding to $\mu_{\max}$, for any real $\phi_k$ \cite[Pg. 550]{Meyer:00}.  
Thus,  
\begin{equation}
\hspace{-0.2cm} \sum\limits_{k=1}^{\lenY} \left( \frac{ \Avec_k^H \CovMat{\myU}^* \Avec_k} {\|\Avec_k\|}\right)^2\!\|\Avec_k\|^2 
 \leq \mu_{\max}^2  \sum\limits_{k=1}^{\lenY}\|\Avec_k\|^2 \leq \mu_{\max}^2 P.
 \label{eqn:UpBound1}
\end{equation}
It follows from \eqref{eqn:UpBound1} that any selection of $\{\Avec_k\}_{k=1}^{\lenY}$ such that $\Avec_k = \left( \myDetVec{c}\right)_k  \myVec{v}_{\max}^*$ and $\sum\limits_{k=1}^{\lenY}\left|\left( \myDetVec{c}\right)_k\right|^2 = \myP$ solves \eqref{eqn:MIProbDef3}. 
As $\CovMat{\myU}^*$ is Hermitian positive semi-definite, it follows that $\mu_{\max}$ is also the largest eigenvalue of $\CovMat{\myU}^*$, and that its corresponding eigenvector is $\myVec{v}_{\max}$, thus proving the theorem.
%
\qed

\vspace{-0.15cm}
\subsection{Proof of Corollary  \ref{pro:waterfil1}}
\label{app:proofPro1}
In order to prove the corollary we show that if the \ac{mmse} matrix is replaced by the \ac{lmmse} matrix ${\myDetMat{E}}_L\big(\tilde{\myA} \big)$, then $\tilde{\myA}\subopt$ in \eqref{eqn:waterfil1} satisfies the conditions of Lemma \ref{lem:optTildeA1}, namely,  $\myDetMat{V}_{\myKron}$ diagonalizes ${\myDetMat{E}}_L\big(\tilde{\myA}\subopt \big)$ and  $\DaMat$ satisfies \eqref{eqn:DmatCond}.  

Using \eqref{eqn:waterfil1} it follows that ${\myDetMat{E}}_L\big(\tilde{\myA}\subopt \big)$ is given by
\ifsingle 
	\begin{align}
	{\myDetMat{E}}_L\big(\tilde{\myA}\subopt \big) 
	&\!=\! \CovMat{\myKron}  \! - \! 
	\CovMat{\myKron} \myDetMat{V}_{\myKron} \DaMat^T  \left(2\SigW\myI_{\lenY}  \! + \! \DaMat \myDetMat{V}_{\myKron}^H \CovMat{\myKron}\myDetMat{V}_{\myKron}\DaMat^T \right)^{ \! - \!1}  \DaMat \myDetMat{V}_{\myKron}^H \CovMat{\myKron} \notag \\
	&\!=\! \CovMat{\myKron}  \! - \! 
	\CovMat{\myKron} \myDetMat{V}_{\myKron} \DaMat^T  \left(2\SigW\myI_{\lenY}  \! + \! \DaMat \myDetMat{D}_{\myKron}\DaMat^T  \right)^{ \! - \!1}  \DaMat \myDetMat{V}_{\myKron}^H \CovMat{\myKron}.
	\label{eqn:proofW1}
	\end{align}
\else
	\begin{align}
	{\myDetMat{E}}_L\big(\tilde{\myA}\subopt \big) 
	&\!=\! \CovMat{\myKron}  \! - \! 
	\CovMat{\myKron} \myDetMat{V}_{\myKron} \DaMat^T  \left(2\SigW\myI_{\lenY}  \! + \! \DaMat \myDetMat{V}_{\myKron}^H \CovMat{\myKron}\myDetMat{V}_{\myKron}\DaMat^T \right)^{ \! - \!1} \notag \\
	&\qquad \qquad \qquad\qquad \qquad \cdot \DaMat \myDetMat{V}_{\myKron}^H \CovMat{\myKron} \notag \\
	&\!=\! \CovMat{\myKron}  \! - \! 
	\CovMat{\myKron} \myDetMat{V}_{\myKron} \DaMat^T  \left(2\SigW\myI_{\lenY}  \! + \! \DaMat \myDetMat{D}_{\myKron}\DaMat^T  \right)^{ \! - \!1} \notag \\
	&\qquad \qquad \qquad\qquad \qquad \cdot \DaMat \myDetMat{V}_{\myKron}^H \CovMat{\myKron}.
	\label{eqn:proofW1}
	\end{align}
\fi 
From \eqref{eqn:proofW1} it follows that ${\myMat{E}}_L\big(\tilde{\myA}\subopt \big) $ is diagonalized by $\myMat{V}_{\myKron}$, and  the eigenvalue matrix is the diagonal matrix given by 
\ifsingle 
	\begin{equation}
	\myDetMat{V}_{\myKron}^H{\myDetMat{E}}_L\big(\tilde{\myA}\subopt \big) \myDetMat{V}_{\myKron}
	= \myDetMat{D}_{\myKron} - 
	\myDetMat{D}_{\myKron} \DaMat^T  \left(2\SigW\myI_{\lenY} + \DaMat \myDetMat{D}_{\myKron}  \DaMat^T \right)^{-1}\DaMat \myDetMat{D}_{\myKron}. 
	\label{eqn:proofW2}
	\end{equation}
\else
	\begin{align}
	&\myDetMat{V}_{\myKron}^H{\myDetMat{E}}_L\big(\tilde{\myA}\subopt \big) \myDetMat{V}_{\myKron} \notag \\
	&= \myDetMat{D}_{\myKron} - 
	\myDetMat{D}_{\myKron} \DaMat^T  \left(2\SigW\myI_{\lenY} + \DaMat \myDetMat{D}_{\myKron}  \DaMat^T \right)^{-1}\DaMat \myDetMat{D}_{\myKron}. 
	\label{eqn:proofW2}
	\end{align}
\fi 
In order to satisfy \eqref{eqn:DmatCond}, for all $k \in \lenYSet$, $\big(\DaMat \big)_{k,k} $ must be non-negative, and  if  $\big(\DaMat \big)_{k,k} > 0$, then  from \eqref{eqn:proofW2}:
\begin{equation}
\eta 
=  \left( \myDetMat{D}_{\myKron}\right)_{k,k} -  \frac{\left( \myDetMat{D}_{\myKron}\right)_{k,k}^2 \left(\DaMat \right)_{k,k}^2}{2\SigW + \left(\DaMat \right)_{k,k}^2\left( \myDetMat{D}_{\myKron}\right)_{k,k}}.
\label{eqn:proofW3}
\end{equation}
Extracting $\big(\DaMat \big)_{k,k}^2$ from \eqref{eqn:proofW3} and setting $\tilde{\eta} \triangleq \frac{2\SigW}{\eta}$ 
yields \eqref{eqn:waterfil0}, and concludes the proof. 
\qed

\vspace{-0.15cm}
\subsection{Proof of Proposition  \ref{pro:KRP}}
\label{app:proofKRP}
Letting $\Avec_k$ be the $k$-th column of $\myA^T$, $k \in \lenYSet$, we note that 
\begin{equation}
\label{eqn:proofKRP1}
\|\myMat{V}\tilde{\myA}\subopt - \myS_{\lenY}\left(\myA \otimes \myA^* \right)\|^2 = \sum\limits_{k=1}^{\lenY} \|\tilde{\Avec}_k\subopt - \Avec_k \otimes \Avec_k^* \|^2.
\end{equation}
Therefore, the solution to the nearest row-wise \ac{krp} problem \eqref{eqn:KRPprobDef} is given by the solutions to the $\lenY$ nearest Kronecker product problems, i.e., for any $k \in \lenYSet$,
\begin{align}
\hat{\Avec}_k\nkrp 
 &= \mathop {{\rm{argmin}}}\limits_{{ \Avec_k} \in \mySet{C}^{\lenU}}\left\|\tilde{\Avec}_k\subopt - \Avec_k \otimes \Avec_k^* \right\|^2 \notag \\
  &\stackrel{(a)}{=} \mathop {{\rm{argmin}}}\limits_{{ \Avec_k} \in \mySet{C}^{\lenU}}\left\|{\rm vec}_\lenU^{-1}\left( \tilde{\Avec}_k\subopt\right)  - \Avec_k^*\Avec_k^T  \right\|^2,
 \label{eqn:KRPprobDef2}
\end{align}
 where $(a)$ follows from \eqref{eqn:KronProp1}.
 %
Solving \eqref{eqn:KRPprobDef2} is facilitated by the following Lemma:
 \begin{lemma}
 	\label{lem:OptProb}
 	For an $\lenU \times \lenU$ matrix $\myDetMat{X}$ with Hermitian part $\myDetMat{M}_X$, 
 	it holds that
 	\begin{equation}
 	\label{eqn:OptProb}
 	\mathop{\rm argmin}\limits_{\myDetVec{v} \in \mySet{C}^\lenU} \left\| \myDetMat{X}-  \myDetVec{v}^*\myDetVec{v}^T \right\|^2 = 	\mathop{\rm argmin}\limits_{\myDetVec{v} \in \mySet{C}^\lenU} \left\| \myDetMat{M}_X-  \myDetVec{v}^*\myDetVec{v}^T \right\|^2.
 	\end{equation}
 \end{lemma}
 {\em Proof:}
 We note that since  $\left\| \myDetMat{B} \right\|^2 = {\rm Tr}\left(\myDetMat{B}\myDetMat{B}^H \right)$, then  
%
%
 \begin{align}
 \left\| \myDetMat{X}-  \myDetVec{v}^*\myDetVec{v}^T \right\|^2 
 &= \left\| \myDetMat{X} \right\|^2 + \left\|  \myDetVec{v}^*\myDetVec{v}^T \right\|^2 -  \myDetVec{v}^T \left( \myDetMat{X} +\myDetMat{X}^H\right)  \myDetVec{v}^* \notag \\
 &\stackrel{(a)}{=} \left\| \myDetMat{X} \right\|^2 + \left\|  \myDetVec{v}^*\myDetVec{v}^T \right\|^2 - 2\myDetVec{v}^T \myDetMat{M}_X \myDetVec{v}^*,
 \label{eqn:OptProb2}
 \end{align}
 where $(a)$ follows since $\myDetMat{M}_X = \frac{1}{2}\big(\myDetMat{X} + \myDetMat{X}^H\big)$.
 Applying the $\mathop{\rm argmin}$ operation to \eqref{eqn:OptProb2} proves the lemma. 
 \qed

From Lemma \ref{lem:OptProb} it follows that  \eqref{eqn:KRPprobDef2} is equivalent to 
\begin{align}
\hat{\Avec}_k\nkrp 
&= \mathop {{\rm{argmin}}}\limits_{{ \Avec_k} \in \mySet{C}^{\lenU}}\|\tilde{\myDetMat{M}}_{k}^{(H)} - \Avec_k^*\Avec_k^T  \|^2,
\label{eqn:KRPprobDef3}  \\
&= \mathop {{\rm{argmin}}}\limits_{{ \Avec_k} \in \mySet{C}^{\lenU}} \left( \| \tilde{\myDetMat{M}}_{k}^{(H)} \|^2 +  \| \Avec_k^*\Avec_k^T  \|^2 -2 \Avec_k^T\tilde{\myDetMat{M}}_{k}^{(H)} \Avec_k^*\right)  \notag \\
&\stackrel{(a)}{=} \mathop {{\rm{argmin}}}\limits_{{ \Avec_k} \in \mySet{C}^{\lenU}}\left(  \| \Avec_k^*\Avec_k^T  \|^2 -2 \Avec_k^H\left( \tilde{\myDetMat{M}}_{k}^{(H)}\right)^*\Avec_k\right) ,
\label{eqn:KRPprobDef3a}
\end{align}
where $(a)$ follows since $\tilde{\myDetMat{M}}_{k}^{(H)}$ does not depend on $\Avec_k$, and since $\Avec_k^T\tilde{\myDetMat{M}}_{k}^{(H)} \Avec_k^*$ is real valued \cite[Pg. 549]{Meyer:00}. 
Since the rank one Hermitian matrix $\Avec_k^*\Avec_k^T$ is positive semi-definite, the Eckart-Young theorem \cite[Thm. 2.4.8]{Golub:13} cannot be used to solve \eqref{eqn:KRPprobDef3}. Consequently, we compute the gradient  of the right hand side of \eqref{eqn:KRPprobDef3a} w.r.t. $\Avec_k$  and set it to zero. This results in
\begin{equation}
\label{eqn:KRPprobDef4}
2\left\|\Avec_k \right\|^2 \Avec_k- 2 \left( \tilde{\myDetMat{M}}_{k}^{(H)}\right)^*\Avec_k = 0.
\end{equation}
In order to satisfy \eqref{eqn:KRPprobDef4}, $\hat{\Avec}_k\nkrp $ must be either the zero vector or an eigenvector of the Hermitian matrix $\big( \tilde{\myDetMat{M}}_{k}^{(H)}\big)^*$ with a non-negative eigenvalue. Specifically, for any non-negative eigenvalue  $\tilde{\mu}_k^p$ of $\tilde{\myDetMat{M}}_{k}^{(H)}$ and  its corresponding unit-norm eigenvector $\tilde{\myVec{v}}_k^p$, we have that  $\big(\tilde{\myVec{v}}_k^p\big)^*$ is an eigenvector of $\big( \tilde{\myDetMat{M}}_{k}^{(H)}\big)^*$ with eigenvalue  $\tilde{\mu}_k^p$, and thus
\eqref{eqn:KRPprobDef4} is satisfied by   $\Avec_k^p =  \sqrt{\tilde{\mu}_k^p}\cdot \big(\tilde{\myVec{v}}_k^p\big)^*$, $p \in \lenUSet$. In order to select the eigenvalue-eigenvector pair which minimizes the Frobenious norm, we plug $\Avec_k^p$ into the right hand side of \eqref{eqn:KRPprobDef3a}, which results in 
\ifsingle 
	\begin{equation}
	\left\|\Avec_k^p \right\|^4 - 2   \left( \Avec_k^p\right) ^H \left( \tilde{\myDetMat{M}}_{k}^{(H)}\right)^*\Avec_k^p  = \left(  \tilde{\mu}_k^p\right) ^2 -2\left(  \tilde{\mu}_k^p\right) ^2
	= - \left(  \tilde{\mu}_k^p\right) ^2.
	\label{eqn:KRPprobDef5}
	\end{equation}
\else
	\begin{align}
	&\left\|\Avec_k^p \right\|^4 - 2   \left( \Avec_k^p\right) ^H \left( \tilde{\myDetMat{M}}_{k}^{(H)}\right)^*\Avec_k^p  \notag \\
	&\qquad = \left(  \tilde{\mu}_k^p\right) ^2 -2\left(  \tilde{\mu}_k^p\right) ^2
	= - \left(  \tilde{\mu}_k^p\right) ^2.
	\label{eqn:KRPprobDef5}
	\end{align}
\fi 
Note that \eqref{eqn:KRPprobDef5} is minimized by the largest eigenvalue. Thus, when some eigenvalues are non-negative then the expression \eqref{eqn:KRPprobDef3a} is minimized by taking the largest non-negative eigenvalue. 
When all the eigenvalues are negative,  $\big(\tilde{\myDetMat{M}}_{k}^{(H)}\big)^*$ is negative definite. In this case, the expression in \eqref{eqn:KRPprobDef3a} is strictly non-negative, hence its minimal value is obtained by setting  $\Avec_k$ to be the all-zero vector.
Consequently, $\hat{\Avec}_k\nkrp  = \sqrt{\max\left(\tilde{\mu}_{k,\max},0\right)} \cdot\tilde{\myVec{v}}_{k,\max}^*$.
%
%
%
\qed

\vspace{-0.15cm}
\subsection{Proof of Proposition  \ref{pro:KRP2}}
\label{app:proofKRP2}
Let  $\Avec_{q}$ be the $q$-th column of $\myA^T$, and recall that $\lenY= b \cdot \lenU$. When $\myA$ corresponds to a masked Fourier measurement matrix \eqref{eqn:KRP2AMatForm} we have that 
the right hand side of \eqref{eqn:KRPprobDef3a}, which results in 
\ifsingle 
	\begin{align}
	\|\myMat{V}\tilde{\myA}\subopt - \myS_{\lenY}\left(\myA \otimes \myA^* \right)\|^2 
	&= \sum\limits_{l=1}^{b}\sum\limits_{k=1}^{\lenU} \|\tilde{\Avec}_{(l-1)\lenU + k}\subopt - \Avec_{(k-1)\lenU + p} \otimes \Avec_{(l-1)\lenU + k}^* \|^2  \notag \\
	&\stackrel{(a)}{=} \sum\limits_{l=1}^{b}\sum\limits_{k=1}^{\lenU}\left\|{\rm vec}_\lenU^{-1}\left( \tilde{\Avec}_{(l-1)\lenU + k}\subopt\right)  - \Avec_{(k-1)\lenU + p}^*\Avec_{(l-1)\lenU + k}^T  \right\|^2 \notag \\
	&\stackrel{(b)}{=} \sum\limits_{l=1}^{b}\sum\limits_{k=1}^{\lenU} \left\|{\rm vec}_\lenU^{-1}\left( \tilde{\Avec}_{(l-1)\lenU + k}\subopt\right)  -  \tFmat{k}^* \gvec_l^* \gvec_l^T \tFmat{k}  \right\|^2,
	\label{eqn:proof2KRP1}
	\end{align}
\else
	\begin{align}
	&\|\myMat{V}\tilde{\myA}\subopt - \myS_{\lenY}\left(\myA \otimes \myA^* \right)\|^2 \notag \\
	&= \sum\limits_{l=1}^{b}\sum\limits_{k=1}^{\lenU} \|\tilde{\Avec}_{(l-1)\lenU + k}\subopt - \Avec_{(k-1)\lenU + p} \otimes \Avec_{(l-1)\lenU + k}^* \|^2  \notag \\
	&\stackrel{(a)}{=} \sum\limits_{l=1}^{b}\sum\limits_{k=1}^{\lenU}\left\|{\rm vec}_\lenU^{-1}\left( \tilde{\Avec}_{(l-1)\lenU + k}\subopt\right)  - \Avec_{(k-1)\lenU + p}^*\Avec_{(l-1)\lenU + k}^T  \right\|^2 \notag \\
	&\stackrel{(b)}{=} \sum\limits_{l=1}^{b}\sum\limits_{k=1}^{\lenU} \left\|{\rm vec}_\lenU^{-1}\left( \tilde{\Avec}_{(l-1)\lenU + k}\subopt\right)  -  \tFmat{k}^* \gvec_l^* \gvec_l^T \tFmat{k}  \right\|^2,
	\label{eqn:proof2KRP1}
	\end{align}
\fi 
where $(a)$ follows from \eqref{eqn:KronProp1};
$(b)$ follows from \eqref{eqn:KRP2AMatForm} since $\Avec_{(l-1)\lenU + k} = \tFmat{k} \gvec_l$. 
From \eqref{eqn:proof2KRP1}, in order to minimize the Frobenious norm, the mask vectors $\gvec_l\maskedF$ should  satisfy 
\begin{equation}
\label{eqn:proof2KRP2}
\!\!\gvec_l\maskedF \!\!=\! \mathop{\rm argmin}\limits_{\gvec_l \in \mySet{C}^\lenU} \sum\limits_{k=1}^{\lenU} \left\|{\rm vec}_\lenU^{-1}\!\left( \tilde{\Avec}_{(l\!-\!1)\lenU\! +\! k}\subopt\right)  \!-\!  \tFmat{k}^* \gvec_l^* \gvec_l^T \tFmat{k}  \right\|^2.
\end{equation} 

As $\tilde{\myDetMat{M}}_{{(l-1)\lenU+k}}^{(H)}$ is the Hermitian part of ${\rm vec}_\lenU^{-1}\big( \tilde{\Avec}_{(l-1)\lenU + k}\subopt\big)$, it follows from Lemma \ref{lem:OptProb} and \eqref{eqn:proof2KRP2} that $\gvec_l\maskedF $ can be obtained from 
\ifsingle 
	\begin{align}
	\gvec_l\maskedF 
	&\!=\! \mathop{\rm argmin}\limits_{\gvec_l \in \mySet{C}^\lenU} \sum\limits_{k=1}^{\lenU} \left\| \tilde{\myDetMat{M}}_{{(l\! - \!1)\lenU\! + \!k}}^{(H)}  \! - \!  \tFmat{k}^* \gvec_l^* \gvec_l^T \tFmat{k}  \right\|^2  \notag \\
	&\!\stackrel{(a)}{=}\!   \mathop{\rm argmin}\limits_{\gvec_l \in \mySet{C}^\lenU} \sum\limits_{k=1}^{\lenU} \left\| \tFmat{k}^* \gvec_l^* \gvec_l^T \tFmat{k}  \right\|^2 - \! 2  \gvec_l^H \tFmat{k}^*  \left( \tilde{\myDetMat{M}}_{{(l\! - \!1)\lenU\! + \!k}}^{(H)}\right) ^*\! \tFmat{k} \gvec_l, 
	\label{eqn:proof2KRP3}
	\end{align} 
\else
	\begin{align}
	\gvec_l\maskedF 
	&\!=\! \mathop{\rm argmin}\limits_{\gvec_l \in \mySet{C}^\lenU} \sum\limits_{k=1}^{\lenU} \left\| \tilde{\myDetMat{M}}_{{(l\! - \!1)\lenU\! + \!k}}^{(H)}  \! - \!  \tFmat{k}^* \gvec_l^* \gvec_l^T \tFmat{k}  \right\|^2  \notag \\
	&\!\stackrel{(a)}{=}\!   \mathop{\rm argmin}\limits_{\gvec_l \in \mySet{C}^\lenU} \sum\limits_{k=1}^{\lenU} \left\| \tFmat{k}^* \gvec_l^* \gvec_l^T \tFmat{k}  \right\|^2 \notag \\ 
	&\qquad \qquad \qquad - \! 2  \gvec_l^H \tFmat{k}^*  \left( \tilde{\myDetMat{M}}_{{(l\! - \!1)\lenU\! + \!k}}^{(H)}\right) ^*\! \tFmat{k} \gvec_l, 
	\label{eqn:proof2KRP3}
	\end{align} 
\fi 
where $(a)$ follows from the same arguments as those leading to \eqref{eqn:KRPprobDef3a}.
Next, we recall that the diagonal elements of $\tFmat{k}$ are in fact the $k$-th row of $\Fmat{\lenU}$, hence  $\tFmat{k}\tFmat{k}^* = \frac{1}{\lenU}\myI_{\lenU}$. Therefore,
\begin{align*}
 \left\| \tFmat{k}^* \gvec_l^* \gvec_l^T \tFmat{k}  \right\|^2 
 &= {\rm Tr}\left(\tFmat{k}^* \gvec_l^* \gvec_l^T \tFmat{k}\tFmat{k}^* \gvec_l^* \gvec_l^T \tFmat{k}  \right) \notag \\
 &=  {\rm Tr}\left( \gvec_l^T \tFmat{k}\tFmat{k}^* \gvec_l^* \gvec_l^T \tFmat{k}\tFmat{k}^* \gvec_l^*  \right) 
 = \frac{1}{\lenU^2}\left\|\gvec_l \right\|^4.
\end{align*}
Plugging this into \eqref{eqn:proof2KRP3} yields
\begin{align}
\gvec_l\maskedF  
&\! = \! \mathop{\rm argmin}\limits_{\gvec_l \in \mySet{C}^\lenU}  \sum\limits_{k\! = \!1}^{\lenU} \frac{1}{\lenU^2}\left\|\gvec_l \right\|^4 \! - \! 2  \sum\limits_{k\! = \!1}^{\lenU} \gvec_l^H \tFmat{k}^*  \left( \tilde{\myDetMat{M}}_{{(l\! - \!1)\lenU\! + \!k}}^{(H)} \right)^*\! \tFmat{k}\gvec_l  \notag \\
&\hspace{-0.2cm}\! = \! \mathop{\rm argmin}\limits_{\gvec_l \in \mySet{C}^\lenU} \frac{\left\|\gvec_l \right\|^4 }{\lenU}\! - \! 2   \gvec_l^H \!\left(\sum\limits_{k\! = \!1}^{\lenU} \tFmat{k} \tilde{\myDetMat{M}}_{{(l\! - \!1)\lenU\! + \!k}}^{(H)} \tFmat{k}^* \right)^*  \gvec_l. 
\label{eqn:proof2KRP4}
\end{align}
In order to find the minimizing vector, we compute the gradient of the right hand side of \eqref{eqn:proof2KRP4} with respect to $\gvec_l$  and equate it to zero, which results in
\begin{equation}
\label{eqn:proof2KRP5}
\frac{2}{\lenU}\left\|\gvec_l \right\|^2 \gvec_l - 2  \left(\sum\limits_{k=1}^{\lenU} \tFmat{k}  \tilde{\myDetMat{M}}_{{(l-1)\lenU+k}}^{(H)} \tFmat{k}^* \right)^*  \gvec_l = 0.
\end{equation}

In order to satisfy \eqref{eqn:proof2KRP5}, $\gvec_l\maskedF$ must be an eigenvector of the $\lenU \times \lenU$ Hermitian matrix $\Big(\sum\limits_{k=1}^{\lenU} \tFmat{k} \tilde{\myDetMat{M}}_{{(l-1)\lenU+k}}^{(H)} \tFmat{k}^* \Big)^* $ with a non-negative eigenvalue, and specifically, for any non-negative eigenvalue  $\bar{\mu}_l^p$ of $\sum\limits_{k=1}^{\lenU} \tFmat{k}  \tilde{\myDetMat{M}}_{{(l-1)\lenU+k}}^{(H)} \tFmat{k}^*$ and  its corresponding unit-norm eigenvector $\bar{\myVec{v}}_l^p$, \eqref{eqn:proof2KRP5} is satisfied by   $\gvec_l^p =  \sqrt{\lenU  \bar{\mu}_l^p}\cdot \big(\bar{\myVec{v}}_l^p\big)^*$, $p \in \lenUSet$. In order to characterize the vector $\gvec_l$ which minimizes the Frobenious norm, we plug $\gvec_l^p$ into the right hand side of \eqref{eqn:proof2KRP4}, which results in 
\ifsingle 
	\begin{equation}
	\frac{1}{\lenU}\left\|\gvec_l^p \right\|^4 - 2   \left( \gvec_l^p\right) ^H \left(\sum\limits_{k=1}^{\lenU} \tFmat{k} \tilde{\myDetMat{M}}_{{(l-1)\lenU+k}}^{(H)} \tFmat{k}^*\right)^* \gvec_l^p   = \frac{1}{\lenU}\left( \lenU \bar{\mu}_l^p\right) ^2 -2\lenU\left(  \bar{\mu}_l^p\right) ^2
	= - \lenU\left(  \bar{\mu}_l^p\right) ^2.
	\label{eqn:proof2KRP6}
	\end{equation}
\else
	\begin{align}
	&\frac{1}{\lenU}\left\|\gvec_l^p \right\|^4 - 2   \left( \gvec_l^p\right) ^H \left(\sum\limits_{k=1}^{\lenU} \tFmat{k} \tilde{\myDetMat{M}}_{{(l-1)\lenU+k}}^{(H)} \tFmat{k}^*\right)^* \gvec_l^p  \notag \\
	&\qquad\qquad \qquad = \frac{1}{\lenU}\left( \lenU \bar{\mu}_l^p\right) ^2 -2\lenU\left(  \bar{\mu}_l^p\right) ^2
	= - \lenU\left(  \bar{\mu}_l^p\right) ^2.
	\label{eqn:proof2KRP6}
	\end{align}
\fi 
Note that \eqref{eqn:proof2KRP6} is minimized by the largest eigenvalue.  Thus, when some eigenvalues are non-negative then the expression \eqref{eqn:proof2KRP4} is minimized by taking the largest non-negative eigenvalue. 
When all the eigenvalues are negative, it follows that $\Big(\sum\limits_{k=1}^{\lenU} \tFmat{k} \tilde{\myDetMat{M}}_{{(l-1)\lenU+k}}^{(H)} \tFmat{k}^*\Big)^*$ is negative definite. In this case, the expression in \eqref{eqn:proof2KRP4} is strictly non-negative, hence its minimal value is obtained by setting  $\gvec_l$ to be the all-zero vector.
Consequently, $\gvec_l\maskedF = \sqrt{\lenU \cdot \max\left(\bar{\mu}_{l,\max},0\right)} \cdot \bar{\myVec{v}}_{l,\max}^*$.
\qed

\end{appendix}


\end{document}